\documentclass[aps,pra,twocolumn,showpacs,groupedaddress]{revtex4-1}

\usepackage{amsfonts,mathrsfs,amsmath,amsthm,amssymb}
\usepackage{epsfig,upgreek}
\usepackage{bm}
\usepackage{color}

\makeatletter
\newsavebox{\@brx}
\newcommand{\llangle}[1][]{\savebox{\@brx}{\(\m@th{#1\langle}\)}%
 \mathopen{\copy\@brx\kern-0.5\wd\@brx\usebox{\@brx}}}
\newcommand{\rrangle}[1][]{\savebox{\@brx}{\(\m@th{#1\rangle}\)}%
 \mathclose{\copy\@brx\kern-0.5\wd\@brx\usebox{\@brx}}}
\makeatother

\begin{document}  
\title {\bf 
%Error-Disturbance 
Error-disturbance 
uncertainty relations in Faraday measurements
}

\author{ Le Bin Ho}
\thanks{Electronic address: binho@riec.tohoku.ac.jp}
\affiliation{Research Institute of Electrical Communication, 
Tohoku University, Sendai, 980-8577, Japan}

\author{Keiichi Edamatsu}
\affiliation{Research Institute of Electrical Communication, 
Tohoku University, Sendai, 980-8577, Japan}

\date{\today}

\begin{abstract}
We examine error-disturbance relations 
in the 
%measurements
quantum measurement
of spin systems 
using an atom-light interface scheme.
We model a 
single spin-1/2 
%particle 
system that
interacts with a polarized light meter 
via a Faraday interaction. 
%As a result, both the meter
%and system states are rotated,
%which imprints the system's information
%on the meter throughout the Faraday effect, 
%and likewise,
%the system 
%will be disturbed by the back-action 
%effect of the interaction. 
We formulate the error and disturbance 
of the model and examine the
uncertainty relations.
We found that for 
%classical polarization coherent light, 
the coherent light meter in pure polarization, 
both the 
%square 
error and 
%square 
disturbance 
behave the cyclic oscillations due to
the Faraday rotation 
%and spin rotation effects. 
in both the light and spin polarizations.
%
%Nevertheless, 
We also examine 
%for 
a class of 
polarization squeezed 
%coherent light
light meter, 
where 
we apply the phase-space approximation 
and characterize the role of 
%the squeezing parameter
squeezing.
%for achieving the error and disturbance.
We derive error-disturbance relations 
for these cases and find that 
the Heisenberg-Arthurs-Kelly uncertainty 
%can be 
is
violated while the 
tight Branciard-Ozawa uncertainty always holds.
%
%Furthermore, 
We note that, 
%for weak interaction approximation,
in the limit of weak interaction strength,
%the weak interaction approximation takes place and
the error and disturbance 
%operators obey
become to obey
the unbiasedness condition
and 
hence the Heisenberg-Arthurs-Kelly 
%holds.
relation holds.
%Finally, analysis of the error
%and disturbance in the measurements 
%of an ensemble spin system is also
%signified. 
The work would contribute to
our understanding of 
%the error and disturbance 
%in the 
%spins measurements 
quantum measurement of spin systems
under the atom-light interface framework,
%and the role of the light meter 
%in the achievable of 
%the error-disturbance relations,
and may hold potential 
%for various 
applications in 
%
%quantum estimation theory,
%including 
quantum metrology,
%and 
%quantum state tomography 
%
quantum state estimation and control.
\end{abstract}
%
%
%\pacs{03.65.Ta, 03.65.Aa, 02.50.-r, 03.67.Ac }
\maketitle

\section{Introduction} 
%{\it Introduction.} --- 
Quantum measurements play a crucial 
role in the characterization of physical systems,
which elucidate hidden quantum properties to 
the classical world \cite{wheeler_zurek_2014}. 
Moreover, many measurements come 
with more 
%observables % than just only one 
than just only one observables
that do not commute,
and thus have  an enormous impact
from the fundamental verification such as 
Bell non-locality and entanglement
\cite{RevModPhys.86.419,RevModPhys.81.865,RevModPhys.84.621}, 
quantum steering \cite{RevModPhys.92.015001},
quantum metrology \cite{PhysRevA.87.012107,Steinlechner2013}
to quantum information technologies including 
%quantum information and 
%quantum computation \cite{nielsen_chuang_2010,furusawa_loock_2011},
quantum key distribution \cite{Bedington2017},
quantum dense coding \cite{bouwmeester_ekert_zeilinger_2011,doi:10.1002/qute.201900011,PhysRevLett.69.2881,PhysRevLett.88.047904}, 
%and 
quantum cryptography
\cite{bouwmeester_ekert_zeilinger_2011,RevModPhys.74.145},
and
non-local quantum measurement
\cite{PhysRevLett.90.010402,PhysRevLett.116.070404,Vidil_2021}.

An important intrinsic property of 
quantum measurements is the uncertainty relation
in which %states that 
it is infeasible to measure
incompatible observables with
arbitrary precision.
This is the fundamental restriction 
in the attainable precision 
of quantum measurements.
In the early stage of quantum mechanics, 
Heisenberg \cite{Heisenberg1927}
was first formulated such
an uncertainty relation between
the position measurement and
the disturbance of the momentum 
%through a $\gamma$-experiment
%\cite{Heisenberg1927}. 
%Since then, there are numerous definitions
%and attracting works on this problem.
%Specially, Ozawa has defined the error
%and disturbance and introduced such an
%error-disturbance relation \cite{}
%that claimed to be universally valid.
%Braciare later also provide a more tighter relation
%for the error and disturbance \cite{}. 
that satisfies % the relation %
%\begin{align}{\label{S_edr}}
$\epsilon_{\bm{q}}\eta_{\bm{p}} \approx \hbar /2,$
%\end{align}
%
where $\epsilon_\circ$ and $\eta_\star$ 
represent the {\it root-mean-square} 
error and {\it root-mean-square} disturbance, 
respectively.
% and $\hbar$ is tzthe reduce 
%Plance's constant. 
The study then was paraphrased under 
the form of the standard deviations 
by Kennard \cite{Kennard1927},
Weyl \cite{weyl_1928},
and later Robertson
\cite{PhysRev.34.163}, 
for a general pair of operators $\bm A$ and $\bm B$, reads 
%
%\begin{align}{\label{S_edr}}
$\sigma_{\bm A}\sigma_{\bm B} \ge \mathcal C_{\bm {A},{\bm B}},$
%\end{align}
%
where $\mathcal C_{\bm {A},{\bm B}} 
= |\langle\psi|[\bm A,\bm B]|\psi\rangle|/2$, 
and $\sigma_{\bm\Lambda} = \sqrt{\langle\bm\Lambda^2\rangle 
- \langle\bm\Lambda\rangle^2}$ 
represents the standard deviation of 
$\mathbf\Lambda$, with $\langle \bm\Lambda\rangle 
= \langle\psi| \bm \Lambda|\psi\rangle$ 
is the expectation value for 
a pure quantum state $|\psi\rangle$, 
and $\mathbf\Lambda \equiv \bm A$ or $\bm B$. 
However, this mathematical relation 
in the form of standard derivation 
has no direct connection to the 
limitation on measurements, 
and thus could not cover
Heisenberg's interpretation uncertainty. 
Preferably, Arthurs and Kelly 
\cite{https://doi.org/10.1002/j.1538-7305.1965.tb01684.x}
provided an error-disturbance relation 
and then was generalized to 
\cite{PhysRevLett.60.2447,ISHIKAWA1991257}
\begin{align}{\label{H_edr}}
\epsilon_{\bm A}\eta_{\bm B}
\ge \mathcal C_{\bm{A},{\bm B}}, 
\end{align}
which states that if the measurement of an observable
$\bm A$ with an error $\epsilon_{\bm A}$, 
then it also disturbs an observable $\bm B$
with a disturbance $\eta_{\bm B}$
satisfying such a relation.
So far, it is known that
this relation is 
not universally valid (see, for example, 
Ref.~\cite{RevModPhys.42.358}.) 
%and also below.)  
Hereafter, we call~\eqref{H_edr}
the Heisenberg-Arthurs-Kelly 
uncertainty. 

Ozawa has theoretically derived 
a universal error-disturbance relation 
\cite{PhysRevA.67.042105,OZAWA2004350}
through an indirect measurement
following the von Neumann paradigms
\cite{neumann}. 
The measurement 
consists of an interaction 
between a quantum system and a meter. 
A measurement of $\bm A$ in the system 
was done indirectly via a measurement of $\bm M$ 
in the meter. %, which causes the error of the measurement.
At the same time, this process
affects back to the system,
and thus it disturbs the subsequent measurement of  
observable $\bm B$ in the system. % \cite{OZAWA2004350}. 
According to Ozawa, the error-disturbance relation 
for any input state $|\psi\rangle$ 
is expressed by 
\cite{PhysRevA.67.042105,OZAWA2004350} 
\begin{align}{\label{O_edr}}
\epsilon_{\bm A}\eta_{\bm B}+\epsilon_{\bm A}
\sigma_{\bm B} + \eta_{\bm B}\sigma_{\bm A} 
\ge \mathcal C_{\bm {A},{\bm B}}\;.
\end{align}
This relation has been experimentally 
confirmed recently by using 
a state-preparation method 
\cite{Erhart2012,
PhysRevA.88.022110,
Baek2013,
PhysRevLett.112.020401,
Edamatsu_2016}, 
weak probe method
\cite{OZAWA200511,Lund_2010,
PhysRevLett.109.100404,
PhysRevLett.112.020402}, 
continuous-variable entangled states
\cite{Liu2019,Liu:19},
and others \cite{PhysRevLett.115.030401,
PhysRevLett.117.140402}.

Subsequently, Branciard 
\cite{Branciard6742,
PhysRevA.89.022124},
and Ozawa
 \cite{ozawa2014errordisturbance}
have considered a rigorous relation reads 
\begin{align}{\label{B_edr}}
\epsilon^2_{\bm A}\sigma^2_{\bm B}
+\sigma^2_{\bm A}\eta^2_{\bm B} 
+2\epsilon_{\bm A}\eta_{\bm B}
\sqrt{\sigma^2_{\bm A}\sigma^2_{\bm B}
-\mathcal{C}^2_{\bm {A},{\bm B}}} 
\ge \mathcal{C}^2_{\bm {A},{\bm B}},
\end{align}
that claimed tighter than relation \eqref{O_edr}
and has been experimentally verified 
\cite{PhysRevLett.112.020401,
PhysRevLett.112.020402,
Liu2019,Liu:19}. 
Hereafter, we call \eqref{B_edr}
the Branciard-Ozawa uncertainty. 
Recently, numerous alternative approaches 
have been used to revisit 
the uncertainty relation 
theoretically and experimentally
\cite{Zheng2020,Zheng2017,
lee2020universal,
Renes2017uncertainty,
tajima2019coherencevariance,
PhysRevLett.112.050401,
PhysRevA.94.062110,
PhysRevLett.115.030401,
yuan2019experimental,
PhysRevD.90.045023,
PhysRevLett.113.260401,
PhysRevA.93.052108,
PhysRevA.94.042104,
PhysRevA.102.042226,
%},
%\cite{
PhysRevA.67.022106,
PhysRevA.89.022106,
PhysRevA.88.022107,
PhysRevA.84.042121,
PhysRevLett.111.160405,
RevModPhys.86.1261,
%}. 
%\cite{%PhysRevLett.111.160405,
PhysRevA.89.012129,
%RevModPhys.86.1261}
%\cite{
PhysRevLett.116.160405,
%PhysRevLett.111.160405,
%PhysRevA.89.012129,
%RevModPhys.86.1261,
PhysRevA.95.040101,
%},
%\cite{
PhysRevLett.122.090404,
%},
%\cite{
PhysRevA.96.022137}.

Recently, the Faraday measurements of spin
based on an atom-light interface framework
have been studied actively 
\cite{RevModPhys.82.1041,
Julsgaard_2003,
Colangelo2017, 
PhysRevA.96.063402,
Atature2007,Vasilyev_2012,
doi:10.1063/1.2721380,
Bao2020,
PhysRevLett.124.110503,
PhysRevA.70.052324,
PhysRevLett.102.125301}. 
It has contributed to our understating 
of quantum measurement 
%of spin systems 
and has various applications in
quantum metrology of atomic ensemble
\cite{RevModPhys.90.035005}, 
quantum information processing
\cite{Monroe2002}, 
strongly correlated systems
\cite{Mehboudi_2015},
and many-body systems \cite{Chen2014}.
The 
%resonant 
Faraday effect 
causes the rotation of the polarized 
light 
%by 
via the interaction with  
the spin system and thus
allows indirect measurement of
the spin system 
%via 
through
the polarized light meter. 
Such a measurement 
contains fundamental limits
in the sensitivity caused by 
the quantum nature of light.
Likewise, 
the back-action of the 
polarized light meter 
perturbs the spin state,
which causes disturbance on
the subsequent measurements 
of the spin system.

Recently, 
%the 
uncertainty relation
in the Faraday measurement
%using preparation and postselection
has been studied
%examined so far using 
by examining the relation between
preparation (prediction) and postselection (retrodiction)
{\cite{Bao2020},
where the authors consider the
Weyl-Robertson relation 
for the 
approximate
canonical position and momentum
of the spin 
of atoms.
However, %as we mentioned above, 
%such the precision figure of merit 
the 
ontained
relation cannot be considered as the error-disturbance relation.
Also, the approximate canonical observables used there 
are only applicable 
in the case of weak interaction and
%for an 
unbiased measurements.
%but not for general cases
%because the accuracy is ignored.
%
%As the atom-light interface has grown rapidly  
%recently, it is increasingly important for 
%such an investigation of its uncertainty relations
%under the full definition for the error and disturbance,
%(both precision and accuracy are included).
Thus,
more precise and appropriate analysis of the error-disturbance uncertainty relation 
in Faraday measurement is necessary.

In this paper, we formulate 
an atom-light interface scheme
in the Faraday measurement
and 
%use Ozawa's definition of 
evaluate the error, disturbance and their 
%for evaluating various 
uncertainty relations.
%based on Ozawa's definitions of error and disturbance
%in generalized quantum measurement.
We consider 
%the 
an atom as a single
spin-1/2 particle interacting with a polarized light meter. 
We first consider 
a classical coherent polarized light
as the light meter.
%In this way, 
Without approximation,
we derive the error and disturbance
%and observe the cyclic ossifications
and their trade-off relation
as functions of the interaction strength.
%as a result of the Faraday rotation
%and spin rotation. 
%Similarly, 
Next, we investigate the case of
polarization squeezed light 
using the 
canonical 
phase-space approximation for the light meter, 
where the squeezing parameter
%takes place
is regarded
as 
%a 
one of the parameters that define the
measurement strength. 
%Under this phase-space approximation, 
%the square error will gradually decrease to zeros
%while the square disturbance increases and reach 
%the maximum of two.
We also examine the case of weak interaction strength,
where the error and disturbance satisfy the joint unbiasedness,
i.e., the condition in which the Arthurs-Kelly uncertainty holds.
We further formulate  
the error-disturbance relations 
in these cases 
%above
and provide that 
the Heisenberg-Arthurs-Kelly uncertainty 
\cite{https://doi.org/10.1002/j.1538-7305.1965.tb01684.x}
can be violated 
while the 
tight 
Branciard-Ozawa uncertainty 
for the qubit system
\cite{Branciard6742}
always holds.
%Notably, the 
%Heisenberg-Arthurs-Kelly uncertainty
%holds under weak interaction approximation
%due to the unbiased measurement.
%Further investigation on an 
%ensemble of spins system reveals
%the increase of both the error 
%and disturbance under 
%a large number of spins.
%Thus, 
Our analysis would contribute to
the understanding of 
the effect of error and disturbance 
as well as 
%the uncertainty principle
their uncertainty relations
in the quantum measurement
under the atom-light interface framework.

This paper is organized as follows. 
We introduce the concept of 
the atom-light interface in Sec.~\ref{secii}.
In Sec.~\ref{seciii}, we derive the error
and disturbance under the atom-light interface framework
for 
%both 
classical coherent light meter
and 
%the class of 
polarization squeezed light meter. 
The error-disturbance relations 
are provided in Sec.~\ref{secvi}.
%Finally, an extension to an
%ensemble spin system is given in 
%Sec.~\ref{secv}.
We give a brief summary and outlook 
in Sec.~\ref{secvii}.

\section{Measurement process} {\label{secii}}
%{\it Postselection measurement.} ---
We consider a measurement model 
in which a spin-1/2 system 
interacts with a polarized light meter 
based on the Faraday interaction under 
the standard von Neumann paradigm \cite{neumann}. 
The spin system is a single particle 
charactered by Pauli matrices 
$\bm \sigma_i,$ with $i = x, y, z$,
while the polarized light meter is given by
the Stokes operators $\bm S_i$
\cite{doi:10.1119/1.1976407}.
For light propagating along the 
$z$-direction, we explicitly have
\begin{align}\label{eq:Sto}
\bm S_0 &= \bm a_H^\dagger \bm a_H
+\bm a_V^\dagger \bm a_V
 = \bm n_H + \bm n_V\; ,\\
\bm S_x &= \bm a_H^\dagger \bm a_H
-\bm a_V^\dagger \bm a_V
 = \bm n_H - \bm n_V\; ,\\
\bm S_y &= \bm a_H^\dagger \bm a_V
+ \bm a_H \bm a_V^\dagger\; ,\\
\bm S_z &= -i(\bm a_H^\dagger \bm a_V
-\bm a_H \bm a_V^\dagger)\; ,
\end{align}
where 
$H$ and $V$ stand for the light modes of horizontal and vertical linear polarizations, respectively,
$\bm a_{H,V}$ $(\bm a_{H,V}^\dagger)$ are
the annihilation (creation)
operators in 
the corresponding
%$H$ and $V$  
polarization modes, 
%respectively,
and $\bm n = \bm a^\dagger\bm a$
the photon number 
operator.
The Stokes operators obey 
the angular momentum commutation
relation $[\bm S_x, \bm S_y]
= 2i\bm S_z$, 
and cyclic permutations.

The unitary evolution of the Faraday 
interaction is given by
\begin{align}\label{eq:ue}
\bm U_T = e^{-ig\bm A\otimes \bm S_z},
\end{align}
where $g = \int_0^Tg(t)\ {\rm d}t$ 
is the interaction strength over 
the time interval $T$. 
Here $\bm A$ is the being measured 
observable in the system.
Under such an atom-light interface, 
the polarization state of the light meter 
rotates through the 
Faraday effect by an amount
proportional to $\bm A$, 
and thus allows the 
indirect measurement of $\bm A$.
Likewise, under the back-action effect, 
the system state is rotated around the 
$z$-axis by an amount proportional to $\bm S_z$,
and thus disturbs the system.  

Assume that the spin system is prepared 
in state $|\psi\rangle$ 
and the light meter state is $|\xi\rangle$.
They are initially uncorrelated, so that
$|\Psi\rangle = |\psi\rangle
\otimes|\xi\rangle$.
The unitary operator $\bm U_T$ 
in Eq.~\eqref{eq:ue}
describes the time evolution of the
joint system-meter during 
the interaction time. 
After the interaction, 
the joint state is given by
%\begin{align}\label{eq:jsfi}
$|\Psi'\rangle = \bm U_T|\Psi\rangle,$
%\end{align}
and the measuring expectation
value of an observable $\bm M$ 
in the meter will be
\begin{align}\label{eq:expect}
\notag\langle(\bm I\otimes \bm M) \rangle &= 
\langle \Psi'|(\bm I\otimes \bm M) |\Psi'\rangle \\
&= \langle\Psi|\bm U_T^\dagger
(\bm I\otimes \bm M)
\bm U_T|\Psi\rangle.
\end{align}

Let us choose $\bm M = \bm S_y$,
and  in the Heisenberg picture, 
we consider 
$(\bm I\otimes \bm S_y)_T = 
\bm U_T^\dagger
(\bm I\otimes \bm S_y)_0
\bm U_T$
is the time-dependent operator
after the interaction.
Particularly, 
for $\bm A^2 = \bm I$,
as in Pauli operators,
using the Baker-Campbell-Hausdorff 
(BCH) formula~\cite{Achilles2012}, 
we obtain
(see Appendix~\ref{appA})
\begin{align}\label{eq:sxt}
(\bm I\otimes \bm S_y)_T = 
(\bm I\otimes \bm S_y)_0 \cos(2g)
+ (\bm A\otimes \bm S_x)_0 \sin(2g).
\end{align}
The subscripts $T$ and 0
stand for the time-dependent  
at the time $T$ and 0,
respectively.
Equation~\eqref{eq:sxt} means 
that the Stokes operators rotate 
about the $z$-axis with the angle $2g$, i.e., the Faraday rotation.
The rotation direction is determined by the sign of $\bm A$;
note that the eigenvalues of $\bm A$ are $\pm1$, since $\bm A^2 = \bm I$.  
%
%In a generally designed measurement, 
Then, we measure the expectation
value of the meter 
%$\langle\bm I\otimes\bm S_y\rangle$
$\langle (\bm I\otimes\bm S_y)_T \rangle$
that provides 
the information of an indirect 
measurement performed on the system. 
In our model, the expectation value gives
\begin{align}\label{eq:sxtf}
% \langle ( \bm S_y)_T \rangle_\xi 
\langle (\bm I\otimes \bm S_y)_T \rangle
%_{\psi\otimes\xi} 
=
\langle \bm S_y \rangle_\xi\cos(2g)
+ \langle \bm 
A\rangle_\psi\langle \bm 
S_x\rangle_\xi\sin(2g).
\end{align}
%where, $\langle \cdots \rangle_\xi$ 
%stands for $\langle\xi|\cdots|\xi\rangle$,
%and 
% $\langle \cdots \rangle_\psi$ 
%stands for $\langle\psi|\cdots|\psi\rangle$.
Here and hereafter, the bra-ket symbol 
$\langle \cdots \rangle$ means 
$\langle\Psi|\cdots|\Psi\rangle= 
{\langle}\psi|\langle\xi| \cdots |\psi\rangle|\xi\rangle$ 
%throughout this paper 
%to distinguish with 
whereas
%the single bra-ket symbol 
$\langle \cdots \rangle_\psi$ 
stands for $\langle\psi|\cdots|\psi\rangle$
and $\langle \cdots \rangle_\xi$ 
for $\langle\xi|\cdots|\xi\rangle$.
%respectively.
%
We omit the subscript 0
in the R.H.S without confusion.
Here, the mean value of the 
meter's observable will
shift from the initial value 
by an amount
proportional to the mean value of 
the system's observable
$\langle A \rangle_\psi$
Without loss of generality, 
we 
%usually 
can
choose the initial mean of the 
meter is zero,
i.e.,
$\langle \bm S_y\rangle_\xi = 0$.
We thus can indirectly measure the 
%mean 
value of the system 
operator $\bm A$
via 
a calibrated meter operator 
%$\tilde{\bm S_y} = \bm S_y / 
%\langle \bm S_x\rangle_\xi\sin(2g)$. 
$\bm M_T ={\bigl(\bm I\otimes \bm S_y\bigr)_T}/
{\langle\bm S_x\rangle_\xi\sin(2g)}$.
The calibration is designed so that 
%the mean value of the measurement 
%$\langle \bm M_T \rangle$
$\bm M_T$
is {\it unbiased},
%with respect to $\langle A \rangle_\psi$ 
i.e., 
%$\langle \bm M_T - (\bm A\otimes\bm I)_0\rangle_\xi=0$
%and thus
$\langle \bm M_T - (\bm A\otimes\bm I)_0\rangle=0$
irrespective of $|\psi\rangle$,
given that $\langle \bm S_y\rangle_\xi = 0$.
%We emphasize that
%in practice, 
%one needs to measure 
%${\bm S_y}$, 
%while 
The calibration 
%term
%$\langle \bm S_x\rangle_\xi\sin(2g)$
factor
$1/\langle \bm S_x\rangle_\xi\sin(2g)$
can be determined 
%beforehand,
%at the input.by using,
%such as a reference detector
%\cite{Colangelo2017}. 
independently 
in practical experiments.

In this 
%process
senario, 
to measure $\bm{A}$
of the system 
%at the time 0 
before the interaction, 
we measure 
%$\tilde{\bm S_y}$ 
${\bm S_y}$ 
of the meter
%at time $T$
after the interaction. 
If these two observables are
perfectly correlated in any given 
system
state 
$|\psi\rangle$, 
%of the system, 
%then they would have the same value and 
the measurement is 
said to be 
accurate
\cite{OZAWA200511,OZAWA2006744}. 
However, in general, 
%there would be an error between them because 
%they are not 
they would not be 
perfectly correlated
and thus become inaccurate
because of possible noise and error in the measurement process. 
Moreover, 
%an 
when another
observable $\bm B$ 
in the system is measured after 
the measurement of
$\bm A$, 
it 
%thus 
would be disturbed by the back-action 
effect caused by the prior interaction in 
the 
$\bm A$ measurement.
In the following, we will 
consider the error and disturbance
in 
%such a 
our 
measurement model.

%The equation of motion
%for observable $\bm \sigma_x$
%in the system
%is given by (see App. \ref{appA})
%\begin{align}\label{app:eq:sigmaxf}
%(\bm \sigma_x\otimes \bm I)_T
%= \bm \sigma_x\otimes
%\cos(2g\bm S_z)
%- \bm \sigma_y\otimes
%\sin(2g\bm S_z). 
%\end{align}

\section{Error and disturbance }{\label{seciii}}

%\subsection{Genereal notations}
%In the following, we will consider 
%the error of an $\bm{A}$ measurement in the system
%through an $\bm{M}$ measurement in the meter 
%In the joint system, we denote 
%$\bm{A}_{\rm in} 
%= \bm{A}\otimes\bm{I}, \text{ and } 
%\bm{B}_{\rm in} 
%= \bm{B}\otimes\bm{I},$
%where $\bm{A}$ and $\bm{B}$ 
%are the observables to be measured 
%in the system. We now also define 
%a meter observable $\bm{M}$ in the meter space, 
%such that it becomes $\bm{M}_{\rm in} = 
%\bm{I}\otimes\bm{M}$ in the joint space
%\cite{PhysRevA.67.042105,OZAWA2004350,Ozawa_2005}.

%The interaction is switched on during 
%a time interval $\sigma t$, where the joint system 
%will evolve under the unitary transformation 
%$\bm U$. 
%After the interaction, 
%based on the Heisenberg picture, 
%the observbables 
%$\bm B$ and $\bm M$
%will transform to
%$\bm{B}_{\rm out} = 
%\bm{U}^\dagger
%(\bm{B}\otimes \bm I)\bm{U},$
%and 
%$\bm{M}_{\rm out}= 
%\bm{U}^\dagger(\bm I\otimes\bm{M})\bm{U},$
%respectively.

\subsection{Exact solution for classical coherent light meter}
In the following, we will consider 
the measurements of $\bm A = \bm \sigma_z$ 
and $\bm B = \bm \sigma_x$ 
in a single spin system.
In the joint space, we denote 
\begin{align}\label{eq:aibi}
\bm A_0 = (\bm \sigma_z 
\otimes\bm{I})_0\;, \text{ and } 
\bm B_0 = (\bm \sigma_x\otimes\bm{I})_0,
\end{align}
for the operators at the time 0.
We denote the measurement
operators at time $T$ as
\begin{align}\label{eq:mobo}
\bm M_T = \dfrac{\bigl(\bm I\otimes \bm S_y\bigr)_T}
{\langle\bm S_x\rangle_\xi\ \sin(2g)}
\;, \text{ and } 
\bm B_T = (\bm \sigma_x\otimes\bm{I})_T.
\end{align}
%
%Concretely, we have
We get 
[See Eqs.~(\ref{appeq:MT}, \ref{app:eq:sigmaxfBT}) 
in Appendix~\ref{appB}
]
\begin{align}\label{eq:MTF}
\bm M_T 
&= \dfrac{\bigl(\bm I\otimes \bm S_y\bigr)_0\cot(2g)}
{\langle\bm S_x\rangle_\xi}
+\dfrac{(\bm \sigma_z\otimes \bm S_x)_0}
{\langle\bm S_x\rangle_\xi}\;,\\
\bm B_T
&= 
%\bm \sigma_x\otimes
%\cos(2g\bm S_z)
%- \bm \sigma_y\otimes
%\sin(2g\bm S_z). 
\bigl(
\bm \sigma_x\otimes
\cos(2g\bm S_z)
\bigr)_0
- 
\bigl(
\bm \sigma_y\otimes
\sin(2g\bm S_z)
\bigr)_0 . 
\label{eq:BTF}
\end{align}

%We emphasize, in general, 
%that $[\bm M_T, \bm B_T] \ne 0$,
%in which the error-disturbance relations 
%(\ref{O_edr}, \ref{B_edr}) are inapplicable. 
%Hereafter, in Secs.~(\ref{seciii}-\ref{secv}),
%we assume small $g$ , so that 
%$[\bm M_T, \bm B_T] = 0$.
%We discuss the general case of 
%$[\bm M_T, \bm B_T] \ne 0$ in Sec.~\ref{secvi}.

%
The error 
can be evaluated by 
the error operator 
$\bm N_{\bm \sigma_z}$,
and the disturbance is defined 
through the disturbance operator 
$\bm D_{\bm \sigma_x}$ as follows
\begin{align}\label{eq:ND}
\bm N_{\bm \sigma_z} 
 = \bm M_T - \bm A_0\;, \text{ and } 
\bm D_{\bm \sigma_x} = 
\bm B_T  -\bm B_0\;.
\end{align}
Then, the square error and  
the square disturbance are given by
\cite{PhysRevA.67.042105,OZAWA2004350,Ozawa_2005},
\begin{align}{\label{error_p}} 
\epsilon_{\bm \sigma_z}^2 = \langle 
\bm N_{\bm \sigma_z}^2\rangle\;,
\text{ and } \eta_{\bm \sigma_x}^2 
= \langle \bm D_{\bm \sigma_x}^2\rangle .
\end{align}
%where the bra-ket symbol 
%$\langle \cdots \rangle$ means 
%${\langle}\psi|\langle\xi| \cdots |\psi\rangle|\xi\rangle$ 
%or $\langle\Psi|\cdots|\Psi\rangle$ throughout this paper 
%to distinguished with the single 
%bra-ket symbol $\langle \cdots \rangle_\psi$ 
%for $\langle\psi|\cdots|\psi\rangle$,
%and $\langle \cdots \rangle_\xi$ 
%for $\langle\xi|\cdots|\xi\rangle$,
%respectively.

%Concretely, 
In the following,  
we choose 
the initial system state  
$|\psi\rangle = \frac{1}{\sqrt{2}}(|0\rangle +i |1\rangle)$,
an eigenstate of $\bm \sigma_y$
% in the spin system,
that maximize the R.H.S. of the error-disturbance relations,
Eqs.~\eqref{H_edr}, \eqref{O_edr} and \eqref{B_edr}. 
We also choose 
%and 
the light meter state 
%of being 
to be
a 
%classical
coherent state in the 
%$H$-polarized
horizontal linear polarization:
\begin{align}\label{eq:alH}
|\xi\rangle 
\equiv 
%|\alpha_H\rangle
|\alpha\rangle_H|0\rangle_V
= \exp\bigl(\alpha\bm a^\dagger_H 
-\alpha^*\bm a_H\bigr)
%|0\rangle, 
|0\rangle_H|0\rangle_V, 
\end{align}
where 
%\textcolor{red}{%
%$H$ and $V$ stand for the light modes of horizontal and vertical linear polarizations, respectively,
%}%
%$\alpha$ is
%the complex eigenvalue 
%of the non-Hermitian operator $\bm a_H$,
$|\alpha\rangle$ is the coherent state
with the coherent amplitude $\alpha$,
and $|0\rangle$ is the vacuum state of light.
For this meter state,
$\langle\bm S_x\rangle_\xi=|\alpha|^2$
and
$\langle\bm S_y\rangle_\xi=\langle\bm S_z\rangle_\xi=0$.
%Thus, 
We readily find 
$\langle\bm N_{\sigma_z}\rangle_\xi=0$
and thus
$\langle\bm N_{\sigma_z}\rangle=0$ 
irrespective of $|\psi\rangle$,
i.e.,  
%$\langle\bm M_T\rangle$
$\bm M_T$
is unbiased
as mentioned earlier.
For the disturbance, however,
$\langle\bm D_{\sigma_x}\rangle_\xi \neq 0$ in general, 
since
$\langle\cos(2g\bm S_z)\rangle_\xi \neq \bm I$
even though
$\langle\bm S_z\rangle_\xi=0$
and
$\langle\sin(2g\bm S_z)\rangle_\xi =0$.
The non-zero mean disturbance comes from the noise (fluctuation) in $\bm S_z$,
%around $\langle\bm S_z\rangle_\xi=0$; the noise 
which
randomly rotates the spin system about the $z$-axis
and effectively reduces the $x$-component of the spin. 
This behavior is the imprint of the 
back-action effect 
on the spin system
caused by the light meter,
which
%that 
disturbs (rotates)
the spin system on its 
%Poincar{\'e} 
Bloch
sphere. 

Then, the square error and disturbance read
(see detailed calculation in Appendix~\ref{appB})
\begin{align}
\epsilon_{\bm \sigma_z}^2
 &=\dfrac{1}{|\alpha|^2 \sin^2(2g)},
%\approx
%\dfrac{1}{4\chi^2}, 
\label{eq:err} \\ 
\eta_{\bm \sigma_x}^2 &
%= 2 \big(1 - e^{-2|a|^2\sin^2(g)}\big).
= 2 \big(1 - e^{-2|a|^2\sin^2g}\big).
%&\approx 4\chi^2, 
\label{eq:dis}
\end{align}
%where $\chi = g|\alpha|$ is the 
%measurement strength.
%
%Here, 
%without loss of generality,
%we can consider the weak interaction
%in the atom-light interface,
%i.e., $g \ll 1$, 
%while the measurement strength $\chi$
%can be made arbitrary by setting $\alpha$
%in the initial light coherent state
%$|\alpha_H\rangle$.
%Consequently, $1/|\alpha|^2
%\ll 1/4\chi^2$ and negligible.
%As a result, the 
%Heisenberg-Arthurs-Kelly 
%uncertainty
%\eqref{H_edr}
%is valid, i.e., 
%$\epsilon_{\bm \sigma_z}^2
%\eta_{\bm \sigma_x}^2=1$.
%
%We also confirm that 
%the joint unbiasedness condition 
%is satisfied, i.e.,
%$\langle \bm N_{\bm \sigma_z}\rangle
%= \langle \bm D_{\bm \sigma_x}\rangle
%= 0$ 
%(see App. \ref{appB}),
%which is sufficient 
%for holding the 
%Heisenberg-Arthurs-Kelly uncertainty
%\cite{PhysRevA.67.042105}.
%
Note that,
for the polarized coherent state of light, 
the root-mean-square noise in $\bm S_y$ 
(and also in $\bm S_x$ and $\bm S_z$)
is $|\alpha|$.
This noise is imprinted in $\bm M_T$ 
%by the calibration factor 
as $|\alpha|/\langle \bm S_x\rangle_\xi\sin(2g) = 1/|\alpha|\sin(2g)$,
and thus results in the square error $\epsilon_{\bm \sigma_z}^2$ in Eq.~\eqref{eq:err}.
Also, 
as mentioned above,
the noise
%(root-mean-square noise $|\alpha|$)
in $\bm S_z$ contributes to the disturbance 
in $\bm\sigma_x$
%$\langle\bm D_{\sigma_x}\rangle$
with a bias
and thus results in 
the square disturbance
%Both the fist and second terms in Eq.~\eqref{eq:BTF} contribute to 
%the square disturbance 
$\eta_{\bm \sigma_x}^2$
%and 
given in Eq.~\eqref{eq:dis}.
% (See Appendix~\ref{appB}).

In Fig.~\ref{fig:ed}, 
we show the square error 
$\epsilon_{\bm \sigma_z}^2$
and square disturbance
$\eta_{\bm \sigma_x}^2$
as functions of the interaction strength $g$
for several coherent amplitudes $|\alpha|^2$.
When $g=n\pi/2$ where $n$ is an integer number, 
$\epsilon_{\bm \sigma_z}^2$ diverges 
because no shift of $\bm S_y$ is expected in the meter.
With increasing $g$,
due to the rotation of the 
light polarization 
%on its Poincar{\'e} sphere
that causes a certain amount of shift of $\bm S_y$ in the meter
depending on $\bm\sigma_z$ in the system, 
the square error 
%first is 
gradually decreases as
$\epsilon^2_{\bm \sigma_z} \approx g^{-2}$.
When $g=\pi/4+n\pi/2$,
$\epsilon_{\bm \sigma_z}^2$ reaches its minimum value $1/|\alpha|^2$,
i.e., the minimum square error that can be achieved by the coherent light meter.
%and then
%gradually reduces to the minimum at $g = \pi/4$
%and increases again 
%at $g=\pi/2$.
%The behavior is then repeated  
%when increasing $g$.
%
Likewise, 
the square disturbance
%increases quadratically with 
%the interaction strength
%and reaches the maximum 
%of two and then decreases to zero
%for $g$ varies from 0 to $\pi$.
%
$\eta_{\bm \sigma_x}^2$
exhibits periodic behavior as a function of $g$. 
When $g=n\pi$, 
%where $n$ is an integer number, 
the square disturbance $\eta_{\bm \sigma_x}^2$ vanishes
%because $B_T=B_0$ for any integer values of $\bm S_z$ in Eq.~\eqref{eq:BTF}.
because the spin system is rotated by integer multiples of $2\pi$
for any integer values of $\bm S_z$
and thus returns to its original state.
This phenomenon can be regarded as a kind of {\it quantum revival},
which essentially reflects the discrete nature of the observable, i.e., $\bm S_z$.
%in this case. 
When $g=\pi/2+n\pi$, 
$\eta_{\bm \sigma_x}^2$ reaches its maximum $2 (1 - e^{-2|a|^2})\sim 2$ for large $|\alpha|$.
In this case, the spin system is rotated about the $z$-axis 
by $0$ or $\pi$ at almost even probabilities depending on the even or odd number of $\bm S_z$,
so that the square disturbance becomes approximately $(2^2+0^2)/2=2$.   
This analysis provides us
%the very first impression about 
a complete and accurate insight on
%the error and disturbance 
%in 
%such spin measurements 
the quantum measurement of spin systems
via the Faraday interaction.

%\section{Error-disturbance relations}\label{seciv}
%\subsection{Error-disturbance relations}
%We consider  
%the Heisenberg-Arthurs-Kelly relation
%[the L.H.S of Eq.~\eqref{H_edr}]
%and the Brainciard-Ozawa relation
%[the L.H.S. of Eq.~\eqref{B_edr}],
%which are denoted as HAK and BO, respectively.
%We denote $\mathscr H$, 
%$\mathscr O$, and  $\mathscr B$ for the Heisenberg, 
%Ozawa, and Brainciard error-disturbance relations, 
%respectively.
%
%With our choice of the spin system, we have 
%$\sigma_{\bm \sigma_z} = 1, 
%\sigma_{\bm \sigma_x} = 1,$
%and $\mathcal{C}_{\bm \sigma_z, \bm \sigma_x} = 1$.
%We straightforwardly rewrite these relations as
%\begin{align}
%{\rm HAK} &= \epsilon_{\bm \sigma_z}^2
%\eta_{\bm \sigma_x}^2 \ge 1, \label{eq:rH} \\
%{\rm BO} &= \epsilon^2_{\bm \sigma_z}
%+\eta^2_{\bm \sigma_x} \ge 1. \label{eq:rB}
%\end{align}
%We also consider a tighter Brainciard-Ozawa relation,
%where the condition of $\bm B^2 = \bm I$ 
%is satisfied, here $\bm B = \bm \sigma_x$.
%Following Refs.~\cite{Branciard6742},
%we replace $\eta_{\bm \sigma_x}$ by 
%$\eta_{\bm \sigma_x}\sqrt{1-\frac{\eta^2_{\bm \sigma_x}}{4}}$
%in Eq.~\eqref{eq:rB}
%and recast it as
%\begin{align}{\label{B_edr_t}}
%{\rm BOt} = 
%\epsilon^2_{\bm \sigma_z}
%+\eta^2_{\bm \sigma_x}
%\Bigl(1-\dfrac{\eta^2_{\bm \sigma_x}}{4}\Bigr)
%\ge 1.
%\end{align}

\begin{figure} [t]
\centering
\includegraphics[width=8.6cm]{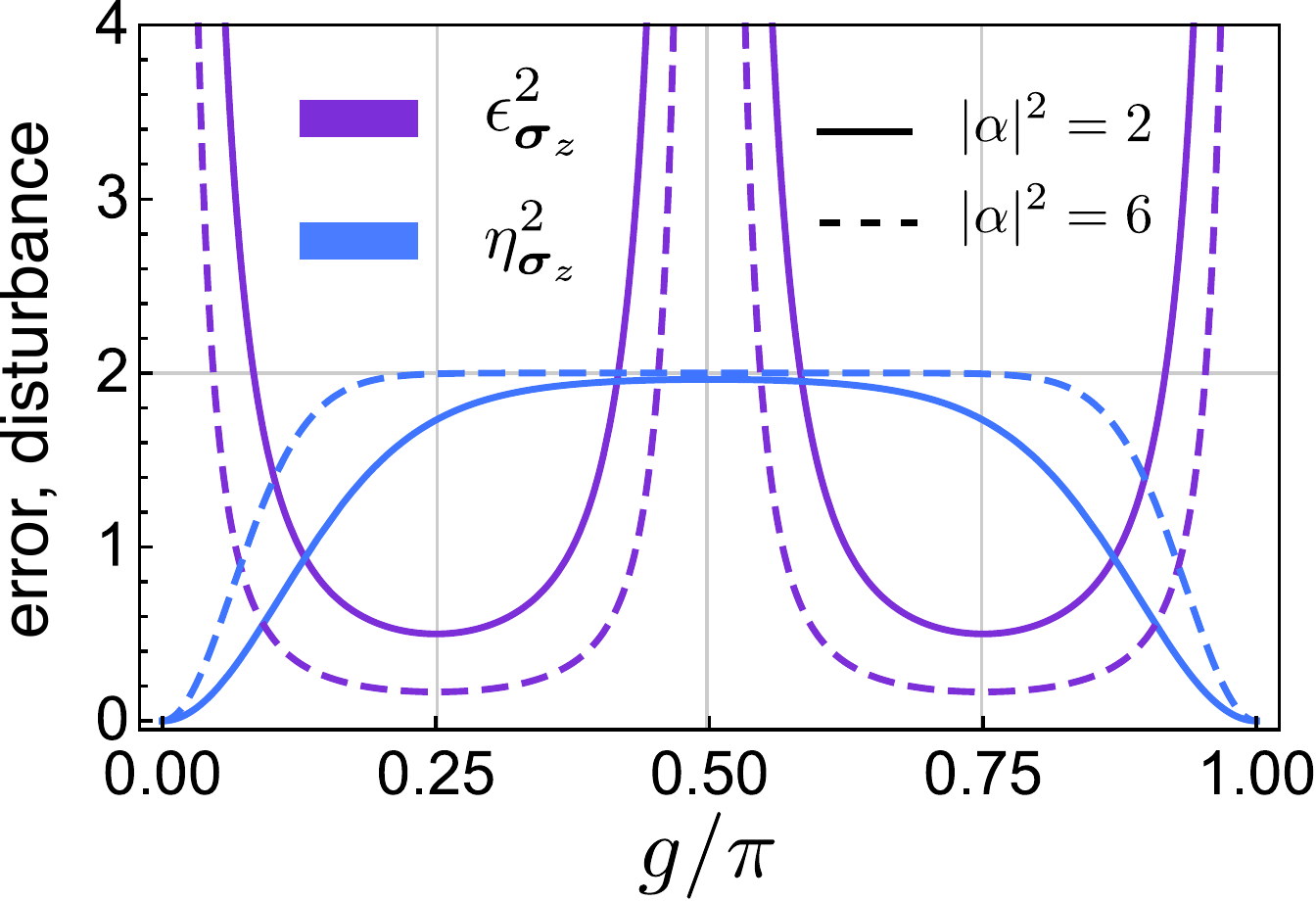}
\caption{
(Color online) 
The plot of the square error 
and square disturbance 
as functions of 
interaction strength $g$ 
for some values of amplitude 
$|\alpha|^2$ as shown in the figure.
The square error is large for small $g$
and reaches the minimum when $g
= \pi/4$ and increases again 
for $g$ increases to
$\pi/2$. The procedure is repeated 
when continuously increasing $g$.
Similarly, the square disturbance 
increases along with $g$ and reaches 
the maximum of two, 
then it reduces to zero as 
$g$ increasing to $\pi$.
}
\label{fig:ed}
\end{figure} 

%\section{Beyond the uncertainty principle}\label{secv}
\subsection{Phase-space 
%approach 
approximation
(canonical approximation) 
for polarization squeezed  light meter}

To further investigate the error and disturbance 
in various light meter states,
we apply the phase-space approximation 
(PSA) for the light system. 
We introduce two canonical operators 
as 
$\bm q \equiv \bm S_y/
%\sqrt{\langle\bm S_x\rangle}$ 
\sqrt{|\langle\bm S_x\rangle|}$ 
and $\bm p \equiv \bm S_z/
\sqrt{|\langle\bm S_x\rangle|}$
for a finite $\langle\bm S_x\rangle$
\cite{Bao2020,RevModPhys.82.1041}.
These operators 
approximately
obey 
the canonical communicator 
relation $[\bm q, \bm p] = 
2i\frac{\bm S_x}
%{|\langle\bm S_x\rangle|}=2i$,
{|\langle\bm S_x\rangle|}\simeq 2i$.
%which is a constant. 
%It is convenient 
%for later calculate the equation of motions
%using the BCH formula, i.e., only its first 
%and second terms are valid 
%while the rest vanish.
%Moreover, using this approximation,
%we also can define various classes of the 
%light meter.  
%%
This approximation is valid 
when $\bm S_x$ can be regarded as a classical positive constant 
that does not change during the measurement process.
Practically, under the PSA
the evolution of $\bm q$ 
in the BCH formula (Eq.~\eqref{app:eq:BCHE} in Appendix A)
is approximated up to its first order (first and second terms). 

Here, 
we discuss the error and disturbance 
using the impact of a class
of the polarization squeezed state 
in the light meter space,
which is given by
\begin{align}\label{eq:xisqueezed}
|\xi\rangle = 
\Bigl(\dfrac{1}{2\pi\sigma^2}\Bigr)^{1/4}
\int e^{-\frac{q^2}{4\sigma^2}} |q\rangle \ {\rm d}q,
\end{align}
where $\sigma$ represents 
the squeezing parameter. 
For $\sigma = 1$, it is a coherent state,
the cases $\sigma < 1$ 
and $\sigma > 1$ correspond 
to an amplitude-squeezed  state
and a phase-squeezed  state,
respectively~\cite{Breitenbach1997}
(See App.~\ref{appC} for detained.)
Here, $q$ and $|q\rangle$ are the eigenvalue 
and eigenstate of the position operator $\bm q$,
such that $\bm q|q\rangle = q|q\rangle$.
We illustrate such a polarization squeezed 
state in a Poincar{\'e} sphere
in Fig.~\ref{fig:poincare}.

\begin{figure} [t]
\centering
\includegraphics[width=6.cm]{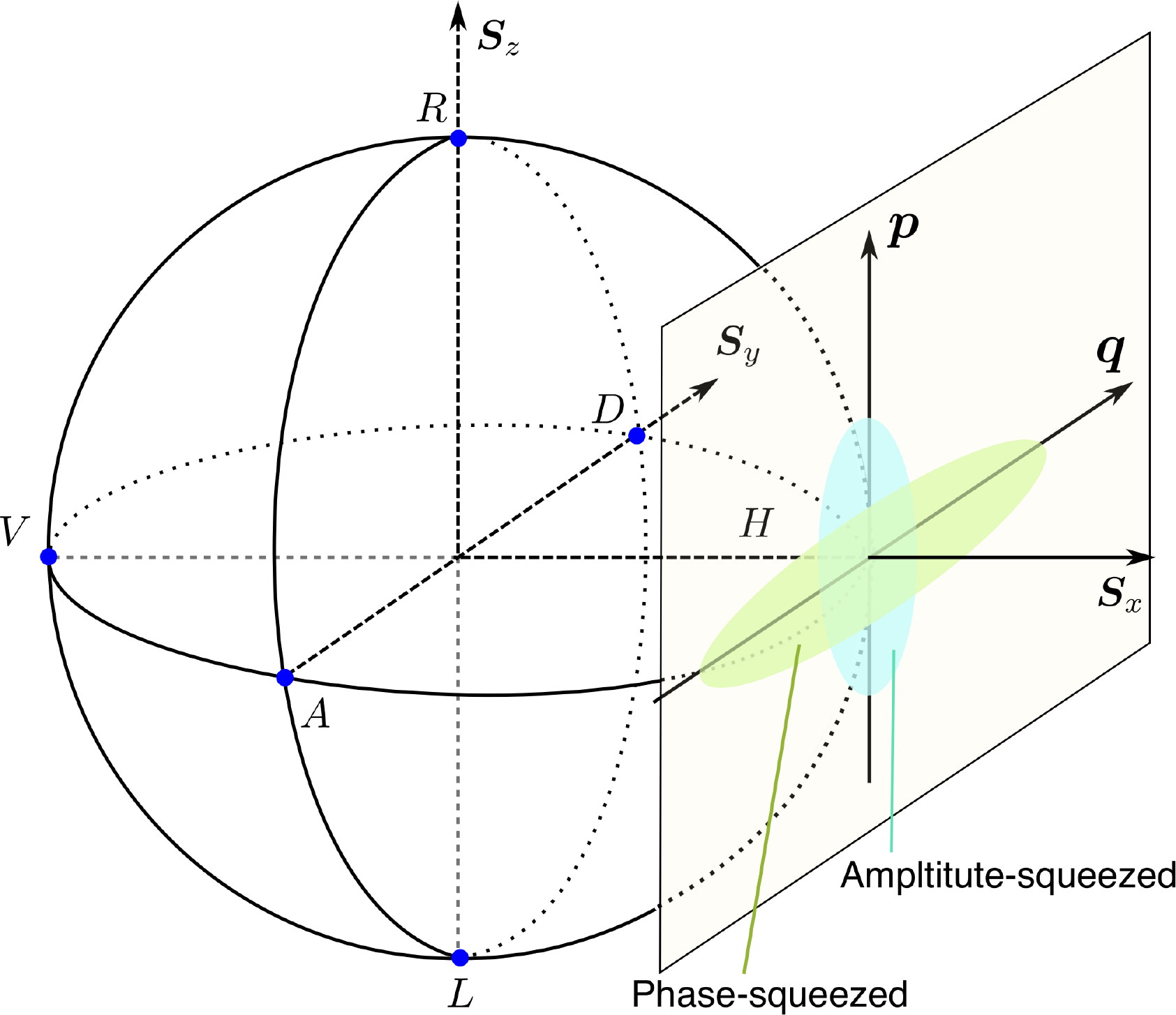}
\caption{
(Color online) Illustration of the 
class of polarization squeezed  
light in the Poincar{\'e} sphere.
}
\label{fig:poincare}
\end{figure} 

The interaction evolution \eqref{eq:ue} is recast as
\begin{align}\label{eq:uere}
\bm U_T = e^{-ig|\alpha|\bm \sigma_z\otimes \bm p},
\end{align}
where we have set $\sqrt{|\langle
\bm S_x\rangle|} = |\alpha|$.
%Similar to above, 
Under the PSA as mentioned above
and using the BCH formula, 
we have
\begin{align}\label{eq:sxtgauss}
(\bm I\otimes \bm q)_T = 
(\bm I\otimes \bm q)_0 
+ 2g|\alpha|(\bm \sigma_z\otimes \bm I)_0.
\end{align}
%Notable that 
%Note that,
%in the PSA,
%we have $[\bm q, \bm p] = 2i$ = constant, 
%thus, only the first and second term in 
%the BCH formula are valid while
%the higher-order terms of 
%$g$ automatically vanish.
%This approximation is equivalent 
%with weak interaction $g$
%for the evolution of the light meter,
%i.e., neglecting the higher-order terms of 
%$g$.
%However, this equivalence is inapplicable 
%for the single spin system since 
%we do not use the phase-space 
%approximation on it.
%

Using
the calibrated meter operator 
$(\bm I\otimes\bm q)_T / 2g|\alpha|$,
we 
%will 
obtain the corresponding 
information of $(\bm \sigma_z\otimes\bm I)_0$
in the system.
Thus,
the error operator is
given by 
%$
\begin{align}
N_{\bm \sigma_z} = (\bm I\otimes\bm q)_T/ 2g|\alpha|
-(\bm \sigma_z\otimes\bm I)_0
= (\bm I\otimes \bm q)_0/ 2g|\alpha| .
\end{align}
%$.
As a result, the square error is appropriate 
to the variance of the 
meter, i.e., 
$\langle \bm q^2\rangle_\xi/ 4g^2|\alpha|^2$
when 
$\langle \bm q\rangle_\xi =0$
. 
Similarly, the square disturbance operator is 
given by 
$2\bigl(1 -
\langle \cos(2g|\alpha|\bm p)\rangle_\xi\bigr)$
%$4g^2\langle\bm p^2\rangle_\xi$
(see App. \ref{appC}).
Straightforward calculating gives
\begin{align}\label{eq:ers}
\epsilon_{\bm \sigma_z}^2
 =\dfrac{1}{4\chi^2}, \
 \text{ and } 
 \eta_{\bm \sigma_x}^2 = 
2\bigl(1 -
e^{-2\chi^2}\bigr),
\end{align}
where $\chi = g|\alpha|/\sigma$
represents the measurement strength. 

In Fig.~\ref{fig:edsq}, we show the square error (solid curve) 
and disturbance (short-dashed curve) 
for PSA
as functions of $\chi$.
%for several values of $g$
When $g|\alpha|$ is fixed (not require to be small), 
hence, the squeezing parameter 
$\sigma$ plays the role of
the measurement strength:
for large $\sigma$ (or small $\chi$)
the measurement is weak,
likewise, for small $\sigma$ (or large $\chi$)
the measurement is strong.
It is natural that for weak measurement,
the square error is large and gradually 
reduces when increasing 
the measurement strength
(solid curve).
Inversely, the square disturbance 
is small for weak measurement 
and gradually increases
when increasing the 
measurement strength 
and reaches the maximum of two
(short-dashed curve).
These results can be explained 
by the ``squeezed" of the meter state: 
%As can be seen from the inset figure, 
larger $\sigma$ means broader Gaussian shape
in the class of the polarization squeezed state,
which is equivalent to ``weak measurement",
while small $\sigma$ refers to the 
narrow of Gaussian shape
which results in a strong measurement. 
Thus, using the class of the 
polarization squeezed state  
will be more convenient 
in some particular cases,
such as using squeezed states 
instead of large photon-number 
coherence states.

\begin{figure} [t]
\centering
\includegraphics[width=8.6cm]{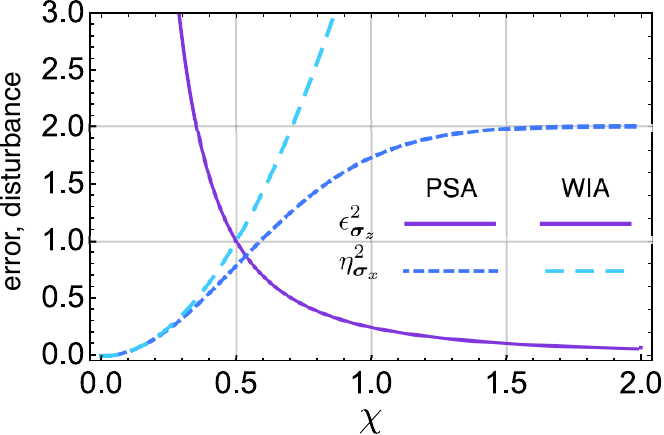}
\caption{
(Color online) Phase-space approximation (SPA):
the plot of square error (solid curve) 
and square disturbance (short-dashed curve)
in the class of squeezed coherent state in the meter
as functions of the measurement strength 
$\chi = g|\alpha|/\sigma$. 
For a fixed of $g|\alpha|$, then $\sigma$ plays the role of
the measurement strength:
for large $\sigma$ the measurement is weak,
for small $\sigma$ the measurement is strong.
Correspondingly, small (large) $\chi$ 
implies weak (strong) measurement. 
Weak interaction approximation (WIA):
the plot of square error (solid curve) 
and square disturbance (long-dashed curve)
%in the weak interaction approach 
in the WIA 
as functions of the measurement strength 
$\chi$.
}
\label{fig:edsq}
\end{figure} 

\subsection{Weak interaction 
%approach
approximation}
In many realistic models, 
the Faraday interaction is used 
in the far off-resonant light 
to avoid completely absorbing 
the classical field \cite{RevModPhys.82.1041}.
For adaptability with such models, 
a weak interaction in the atom-light 
can be made (see, for example, 
Ref.~\cite{Colangelo2017}).

In this subsection, to investigate 
the impact of the error end disturbance
in such a far off-resonant region,
we consider a weak interaction 
%approach 
approximation
(WIA) , i.e., 
%$g \ll 1$.
$\chi \ll 1$.
In this 
%approach,
approximation,
we have 
\begin{align}\label{app:err-app}
\epsilon_{\bm \sigma_z}^2 \approx
\dfrac{1}{4\chi^2},
\text {and }
\eta_{\bm\sigma_x}^2\approx 
4\chi^2,
\end{align}
where $\chi = g|\alpha|$
(see, App.~\ref{appBS}).
Noting that in the 
coherent-state case, $\sigma = 1$.
We show Fig.~\ref{fig:edsq} for
the square error (solid curve) and
square disturbance (long-dashed curve),
denoted by `WIA'.
While the square error is the same as in the 
PSA case, the square disturbance gradually increasing
from zero when increasing $\chi$.
This square disturbance is different from 
that of the PSA case because 
the WIA is applied to both the
spin system and the light meter
(we can neglect the higher-order terms of $g$ 
in both the spin system and the light meter.)
It is thus provides us 
the impact of weak Faraday interaction 
on the error and disturbance 
in spin measurements.
%
%or
%\begin{align}\label{app:err-app2}
%\epsilon_{\bm \sigma_z}^2
% \approx \dfrac{1}{4\chi^2}, \
% \text{ and } 
% \eta_{\bm \sigma_x}^2 \approx 
%4\chi^2,
%\end{align}
%for the phase-space approach.
%
We also confirm that 
the joint unbiasedness condition 
is satisfied, i.e.,
$\langle \bm N_{\bm \sigma_z}\rangle
= \langle \bm D_{\bm \sigma_x}\rangle
= 0$
irrespective of the initial system state
$|\psi\rangle$ 
(see App. \ref{appBS}),
which is sufficient 
for holding the 
Heisenberg-Arthurs-Kelly uncertainty
\cite{PhysRevA.67.042105},
i.e., 
$\epsilon_{\bm \sigma_z}^2
\eta_{\bm \sigma_x}^2=1$.

\section{Error-disturbance relations}\label{secvi}
This section examines the error-disturbance relations
for the measurement of a single spin system 
with two cases of the meter state:
exact solution of 
the classical coherent light [Eq.~\eqref{eq:alH}]
and phase-space approximation 
(PSA) of the polarization 
squeezed light 
[Eq.~\eqref{eq:xisqueezed}.]
We consider  
the Heisenberg-Arthurs-Kelly relation
[the L.H.S of Eq.~\eqref{H_edr}]
and the Brainciard-Ozawa relation
[the L.H.S. of Eq.~\eqref{B_edr}],
which are denoted as HAK and BO, respectively.
%We denote $\mathscr H$, 
%$\mathscr O$, and  $\mathscr B$ for the Heisenberg, 
%Ozawa, and Brainciard error-disturbance relations, 
%respectively.
%
With our choice of the spin system, we have 
$\Delta_{\bm \sigma_z} = 1, 
\Delta_{\bm \sigma_x} = 1,$
and $\mathcal{C}_{\bm \sigma_z, \bm \sigma_x} = 1$.
We straightforwardly rewrite these relations as
\begin{align}
{\rm HAK} &= \epsilon_{\bm \sigma_z}^2
\eta_{\bm \sigma_x}^2 \ge 1, \label{eq:rH} \\
{\rm BO} &= \epsilon^2_{\bm \sigma_z}
+\eta^2_{\bm \sigma_x} \ge 1. \label{eq:rB}
\end{align}
We also consider a tighter Brainciard-Ozawa relation,
where the condition of $\bm B^2 = \bm I$ 
is satisfied, here $\bm B = \bm \sigma_x$.
Following Refs.~\cite{Branciard6742},
we replace $\eta_{\bm \sigma_x}$ by 
$\eta_{\bm \sigma_x}\sqrt{1-\frac{\eta^2_{\bm \sigma_x}}{4}}$
in Eq.~\eqref{eq:rB}
and recast it as
\begin{align}{\label{B_edr_t}}
{\rm BOt} = 
\epsilon^2_{\bm \sigma_z}
+\eta^2_{\bm \sigma_x}
\Bigl(1-\dfrac{\eta^2_{\bm \sigma_x}}{4}\Bigr)
\ge 1.
\end{align}

%The square error and disturbance give
%\begin{align}\label{eq:sqasd}
%\epsilon_{\bm \sigma_z}^2
% = \dfrac{1}{4\widetilde\chi^2},
% \text{ and }
%\eta_{\bm \sigma_x}^2 = 
%2(1-e^{-2\widetilde\chi^2}).
%\end{align}
%For small $g$ approximation, 
%these equations will reduce to those forms as
%shown in Eq.~\eqref{eq:ers}.

%We show the square error and disturbance 
%as functions of  $\widetilde\chi$ in the inset Fig.~\ref{fig:ged}. 
%While the square error will reduces,
%the square disturbance gradually increases 
%and reaches the saturation of two 
%when increasing $\widetilde\chi$.
%
We examine an error-disturbance tradeoff,
which shows the dependence of the disturbance 
on the error or vice versa. The result is shown in 
Fig.~\ref{fig:ged}. It can be seen that the 
error-disturbance tradeoff in the exact case 
behaves the dependence on the Faraday and spin rotations: 
for small $g$, the error is large and the disturbance is small, 
then increasing $g$ results in the reducing of the error and
increasing of the disturbance. 
After the error reaches the minimum, 
the disturbance toward two,
while the error gradually increases,
results in a straight line in the tradeoff.
Here, we show the result for $|\alpha|^2 = 6$.
For large $|\alpha|^2$, 
the tradeoff asymptotically reaches that of
%the phase-space-approach case.
the PSA.
Moreover, the tradeoff can reach 
the BOt relation 
%in the phase-space-approach case, 
in the PSA, 
while the HAK relation is violated.
Concretely, it can be seen that for 
large square error (small $\chi$), 
the error-disturbance tradeoff reaches the HAK bound,
while for small square error (large $\chi$), 
the error-disturbance tradeoff
reaches the maximum of two, 
the BOt bound.
 
\begin{figure} [t]
\centering
\includegraphics[width=8.6cm]{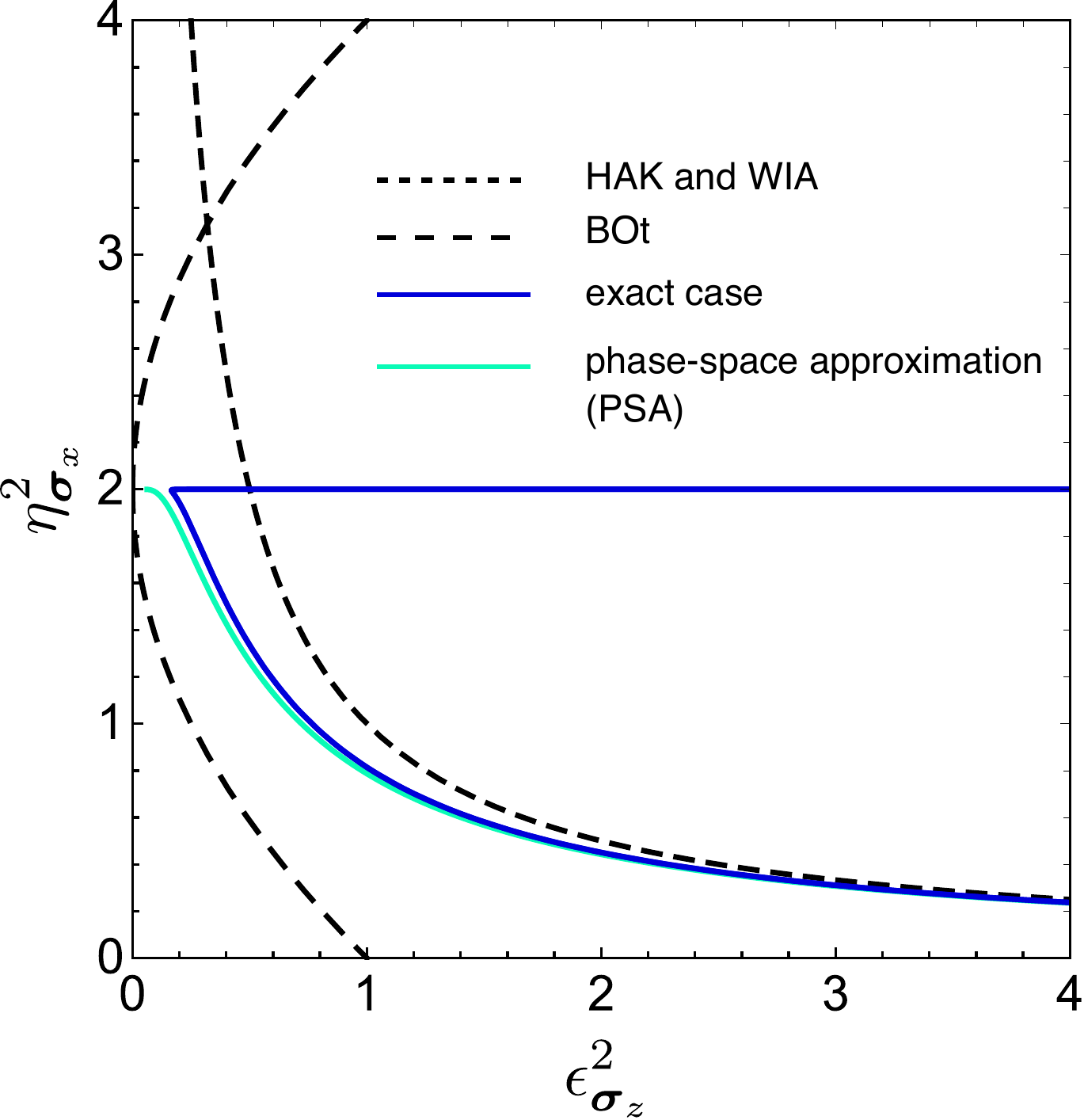}
\caption{
(Color online) 
The error-disturbance tradeoffs. 
The short-dashed curve is the 
Heisenberg-Arthurs-Kelly bound given in 
Eq.~\eqref{eq:rH} and denoted by HAK. 
The left region is the forbidden region 
where the HAK relation is violated. 
Similarly, the long-dashed curve is the
tight Branciard-Ozawa bound 
given in Eq.~\eqref{B_edr_t} and denoted by BOt. 
The solid curves show the error-disturbance tradeoff 
obtained from the atom-light interface 
in this work for two cases of exact solusion
and phase space approximation (PSA).
For the weak interaction approximation (WIA)
it follows the HAK bound. 
}
\label{fig:ged}
\end{figure} 

\section{Conclusion} {\label{secvii}}
%{\it Conclusion.} --- 
We have discussed the 
%error and the disturbance 
error, disturbance and their uncertainly relations 
%on an atom-light interface model. 
%in the framework of atom-light interaction
in Faraday measurements.
%
%Thereby, the measurements of incompatible 
%observables cannot attain arbitrary precision 
%and accuracy due to the fundamental restriction 
%of quantum measurement.
%
%By modeling the error and disturbance 
%under the terms of precision and accuracy,
%we evaluated the uncertainty 
%of quantum measurement 
%in the atom-light interface framework.
%in the framework of atom-light interaction,
For a single spin interacting with coherent polarization of the light meter,
we derived the exact behaviors of error and disturbance without approximation. 
Under the Faraday rotation 
of the coherent light polarization
and its back-action to the spin system,
the error and disturbance behave cyclic oscillations. 
%and the corresponding trade-off relation 
%violates the Heisenberg-Arthurs-Kelly uncertainty.
%Moreover, under the class 
In the case of polarization squeezed light meter,
to which we apply the canonical phase-space approximation,
the squeezing parameter acts as a factor that modifies the measurement strength.
%we map these rotations on 
%the phase-space representation,
%and thus we found that the square error 
%will reduce toward zeros
%while the square disturbance 
%will increase and reach the maximum of two
%when increasing the
%measurement strength.
In this approximation, 
the square error monotonically decreases to 0 
while the square disturbance monotonically increases and approaches to 2 
with increasing measurement strength.
%
%Further investigation on 
%error-disturbance relations 
%where the
In the cases above,
Heisenberg-Arthurs-Kelly uncertainty is violated 
while the tight Branciard-Ozawa uncertainty always holds.
It is worth mentioning that, 
under the weak interaction 
%assumption,
approximation,
the Heisenberg-Arthurs-Kelly 
uncertainty holds 
%due to unbiased measurement.
because the error and disturbance both become to satisfy the unbiasedness.
%Finally, we formulate
%the measurements of 
%the ensemble spin system
%also reveals the increase 
%in both the error and disturbance 
%when increasing the number of spins 
%as a consequence of 
%the spin's correlation and
%quantum fluctuation.

Our analysis would contribute to
%the 
deeper
understanding of 
%the 
error,
%and 
disturbance 
%as well as 
%the 
and 
uncertainty 
%principle
relations
in quantum measurements
under the atom-light interface,
and 
%can 
provide 
%an effective tool for investigating 
an insight into
quantum metrology
\cite{Giovannetti2011,Maga_a_Loaiza_2019},
quantum sensing~\cite{RevModPhys.89.035002},
and quantum state estimation
\cite{paris_2011}.
This analytical work 
%also can be served as 
would also be
a testbed for 
further experimental studies.

\begin{acknowledgments}
This work was supported by 
JSPS KAKENHI Grant Number 20F20021
and by MEXT Quantum Leap Flagship Program 
(MEXT Q-LEAP) Grant Number JPMXS0118067581.
\end{acknowledgments}
%{\it Acknowledgments.} --- 

%\begin{widetext}

\appendix
\setcounter{equation}{0}
\renewcommand{\theequation}{A.\arabic{equation}}
\section{Heisenberg equation of motions}\label{appA}

\subsection{Heisenberg equation for $\bm S_y$}
In the Heisenberg picture, 
the meter's operator $\bm S_y$ 
evolves with time according to
\begin{align}\label{app:eq:Sy}
\bigl(\bm I\otimes \bm S_y\bigr)_T = 
\bm U_T^\dagger
\bigl(\bm I\otimes \bm S_y\bigr)_0
\bm U_T
%.
,
\end{align}
where $\bm U_T = e^{-ig\bm A\otimes\bm S_z}$. 
%Here, $\bm U_T = e^{-ig\bm A\otimes\bm S_z}$. 
%where we consider the case $\bm A^2 = \bm I$,
%such as Pauli 
%%matrices. 
%\textcolor{red}{%
%operators.
%}%

Using the Baker–Campbell–Hausdorff formula
\cite{Achilles2012} 
\begin{align}\label{app:eq:BCHE}
e^{\bm E}\bm Fe^{-\bm E}
= \bm F + [\bm E, \bm F]
+\dfrac{1}{2!}[\bm E, [\bm E, \bm F]] + \cdots,
\end{align}
for $\bm E = ig\bm A\otimes \bm S_z$ 
and $\bm F = \bm I\otimes\bm S_y$,
we have 
\begin{align}
\notag & \bm F = \bm I\otimes\bm S_y,\\
\notag & [\bm E, \bm F] = 
	ig[\bm A\otimes \bm S_z,
	\bm I\otimes \bm S_y]\\
	\notag&\hspace{0.9cm} = 
	ig\bm A\otimes [\bm S_z,\bm S_y]\\
	\notag&\hspace{0.9cm} = 
	2g\bm A\otimes\bm S_x\\
\notag & \dfrac{1}{2!}[\bm E, [\bm E, \bm F]] 
= -\dfrac{(2g)^2}{2!} 
%\bm I
\bm A^2
\otimes\bm S_y,
\\
\notag & \dfrac{1}{3!}[\bm E,[\bm E, [\bm E, \bm F]]]
= -\dfrac{(2g)^3}{3!} 
%\bm A
\bm A^3
\otimes\bm S_x,
\\
\notag &\cdots.
\end{align}
Then Eq.~\eqref{app:eq:BCHE} is recast as
\begin{align}\label{app:eq:BCHEr}
e^{\bm E}\bm Fe^{-\bm E}
= 
%\bigl(\bm I\otimes\bm S_y \bigr)_0 \cos(2g)
%+ \bigl(\bm A\otimes\bm S_x\bigr)_0 \sin(2g). 
\bigl(\cos(2g\bm A)\otimes\bm S_y \bigr)_0 
+ \bigl(\sin(2g\bm A)\otimes\bm S_x\bigr)_0 . 
\end{align}
Submitting Eq.~\eqref{app:eq:BCHEr} 
into the R.H.S. of 
Eq.~\eqref{app:eq:Sy}, we have 
\begin{align}\label{app:eq:Syf0}
(\bm I\otimes \bm S_y)_T
= 
%\bigl(\bm I\otimes\bm S_y \bigr)_0 \cos(2g)
%+ \bigl(\bm A\otimes\bm S_x\bigr)_0 \sin(2g). 
\bigl(\cos(2g\bm A)\otimes\bm S_y \bigr)_0 
+ \bigl(\sin(2g\bm A)\otimes\bm S_x\bigr)_0 . 
\end{align}
In the case of 
$\bm A^2 = \bm I$,
as in Pauli operators,
we obtain
\begin{align}\label{app:eq:Syf}
(\bm I\otimes \bm S_y)_T
= 
\bigl(\bm I\otimes\bm S_y \bigr)_0 \cos(2g)
+ \bigl(\bm A\otimes\bm S_x\bigr)_0 \sin(2g). 
\end{align}
which is given in Eq.~\eqref{eq:sxt}
in the main text.

%Here, $\bm U_T = e^{-ig\bm A\otimes\bm S_z}$.
%%where we consider the case $\bm A^2 = \bm I$,
%%such as Pauli matrices. 
%For small $g$, we expend 
%$\bm U_T \approx \bm I -ig\bm A\otimes\bm S_z$.
%Then, Eq.~\eqref{app:eq:Sy} explicitly gives
% \begin{align}\label{app:eq:Sy_exp}
%\notag(\bm I\otimes \bm S_y)_T &\approx 
%\Bigl(\bm I + ig\bm A\otimes\bm S_z\Bigr)
%\bigl(\bm I\otimes \bm S_y\bigr)_0
%\Bigl(\bm I - ig\bm A\otimes\bm S_z\Bigr)\\
%\notag&=\bigl(\bm I\otimes\bm S_y\bigr)_0 - 
%ig \bm A\otimes 
%\bigl[\bm S_y,\bm S_z\bigr]\\
%&= \bigl(\bm I\otimes\bm S_y\bigr)_0
%+ 2g\bm A\otimes \bm S_x,
%\end{align}
%which is given in Eq.~\eqref{eq:sxt}
%in the main text.

\subsection{Heisenberg equation for $\bm \sigma_x$}

Next, 
we 
consider the particular case where 
$\bm A =\bm\sigma_z$
and
$\bm B =\bm\sigma_x$, 
and
calculate the Heisenberg equation
of motion
for $\bm \sigma_x$ in the spin system. 
%in the same way.
We consider
\begin{align}\label{app:eq:sx}
\bigl(\bm \sigma_x\otimes\bm I\bigr)_T  
= \bm U^\dagger_T 
\bigl(\bm \sigma_x\otimes\bm I\bigr)_0
 \bm U_T.
\end{align}
%Particularly, we consider 
where
$\bm U_T = e^{-ig\bm \sigma_z\otimes\bm S_z}$.
Using the BCH formula
%\cite{Achilles2012} 
%\begin{align}\label{app:eq:BCH}
%e^{\bm E}\bm Fe^{-\bm E}
%= \bm F + [\bm E, \bm F]
%+\dfrac{1}{2!}[\bm E, [\bm E, \bm F]] + \cdots,
%\end{align}
for $\bm E = ig\bm {\bm \sigma}_z\otimes \bm S_z$ 
and $\bm F = \bm \sigma_x\otimes\bm I$,
we have 
\begin{align}
\notag & \bm F =  \bm \sigma_x\otimes\bm I,\\
\notag & [\bm E, \bm F] = ig 
[\bm {\bm \sigma}_z\otimes \bm S_z,
\bm \sigma_x\otimes\bm I]\\
\notag &\hspace{0.95cm} = 
ig[{\bm \sigma}_z,{\bm \sigma}_x]
\otimes \bm S_z\\
\notag &\hspace{0.95cm} = 
-2g{\bm \sigma}_y\otimes \bm S_z,\\
\notag & \dfrac{1}{2!}[\bm E, [\bm E, \bm F]] 
= -\dfrac{(2g)^2}{2!} \bm \sigma_x\otimes\bm S_z^2,\\
\notag &\cdots.
\end{align}
Then Eq.~\eqref{app:eq:sx} gives 
\begin{align}\label{app:eq:sigmaxf}
(\bm \sigma_x\otimes \bm I)_T
= 
%\bm \sigma_x\otimes
%\cos(2g\bm S_z)
%- \bm \sigma_y\otimes
%\sin(2g\bm S_z). 
\bigl(
\bm \sigma_x\otimes
\cos(2g\bm S_z)
\bigr)_0
- 
\bigl(
\bm \sigma_y\otimes
\sin(2g\bm S_z)
\bigr)_0 .
\end{align}

\setcounter{equation}{0}
\renewcommand{\theequation}{B.\arabic{equation}}
\section{Error and disturbance}\label{appB}
In this section, we provice detailed calculation of 
%{\it root mean-square (rms)} 
{\it root mean-square} ({\it rms}) 
error $\epsilon_{\bm \sigma_z}$ 
and 
%{\it root mean-square} 
{\it rms}
disturbance $\eta_{\bm \sigma_x}$. 

\subsection{The error}
%The error can be calculated directly from 
%Eq.~\eqref{eq:err} 
%\begin{align}{\label{app:eq:err}}
%\bm N_{\bm \sigma_z} 
%= \dfrac{1}{\sin2g \langle\bm S_x\rangle_\xi}
%\bigl(\bm I\otimes \bm S_y\bigr)_T
% - \bigl(\bm \sigma_z\otimes\bm I\bigr)_0,
%\end{align}
%in the main text.

We first consider $\bm M_T$
operator:
\begin{align}\label{appeq:MT}
\notag \bm M_T &=
\dfrac{1}{\langle\bm S_x\rangle_\xi \sin(2g)}
\bigl(\bm I\otimes \bm S_y\bigr)_T \\
& = \dfrac{\bigl(\bm I\otimes \bm S_y\bigr)_0 \cot(2g)}
{\langle\bm S_x\rangle_\xi}
+\dfrac{(\bm \sigma_z\otimes \bm S_x)_0}
{\langle\bm S_x\rangle_\xi}\;,
\end{align}
where we have used $\bm A = \bm \sigma_z$
in Eq. \eqref{app:eq:Syf}.
Then, the noise operator is given by
\begin{align}\label{appeq:N}
\bm N_{\bm \sigma_z} 
 =  \dfrac{\bigl(\bm I\otimes \bm S_y\bigr)_0 \cot(2g)}
{\langle\bm S_x\rangle_\xi}
+\dfrac{(\bm \sigma_z\otimes \bm S_x)_0}
{\langle\bm S_x\rangle_\xi}
 - \bigl(\bm \sigma_z\otimes\bm I\bigr)_0\;.
\end{align}
Then, we obtain
\begin{align}{\label{app:eq:N21}}
\bm N_{\bm \sigma_z}^2
= \Bigl[
\underbrace{
       \dfrac{\bigl(\bm I\otimes \bm S_y\bigr)_0 \cot(2g)}
       {\langle\bm S_x\rangle_\xi}  
        }_{\bm Y}
+ \underbrace{
        \dfrac{(\bm \sigma_z\otimes\bm S_x)_0}
        {\langle\bm S_x\rangle_\xi}
        }_{\bm X}
 - \underbrace{
        \bigl(\bm \sigma_z\otimes\bm I\bigr)_0
        }_{\bm Z}
 \Bigr]^2.
\end{align}
Now, we calculate the average 
$\langle\bm N_{\bm \sigma_z}^2\rangle$ 
over the initial joint 
state $|\psi\rangle\otimes|\xi\rangle$.
We have
\begin{align}
\notag & \langle \bm Y^2\rangle = 
\dfrac{\langle \bm S_y^2\rangle_\xi \cot^2(2g)}
{\langle\bm S_x\rangle_\xi^2}\ ;
\langle \bm X^2\rangle 
= \dfrac{\langle\bm S_x^2\rangle_\xi}
{\langle\bm S_x\rangle_\xi^2}\ ;
 \langle \bm Z^2\rangle = 1,\\
\notag & \langle \bm Y\bm X\rangle 
=\langle \bm X\bm Y\rangle = 0,
\langle {\bm Y\bm Z}\rangle = 
\langle {\bm Z\bm Y}\rangle = 0,\\
\notag & \langle {\bm X\bm Z}\rangle = 
\langle {\bm Z\bm X}\rangle = 1.
\end{align}
%As a result, we have 
%\begin{align}\label{app:eq:N2}
%\langle\bm N_{\bm \sigma_z}^2\rangle
%= \cot^22g
%\dfrac{\langle \bm S_y^2\rangle_\xi}
%{\langle
%\bm S_x\rangle_\xi^2}.
%\end{align}
Explicitly, we express the meter coherent state
$|\xi\rangle$ into two modes as
$|\xi\rangle = |\alpha_H, 0_V\rangle$.
We have
\begin{align}\label{app:eq:Sx}
\langle \bm S_x\rangle_\xi =
\langle\alpha_H,0_V|
\bigl(\bm a_H^\dagger\bm a_H
-\bm a_V^\dagger\bm a_V\bigr)
|\alpha_H,0_V\rangle
=|\alpha|^2,
\end{align}
\begin{align}\label{app:eq:Sy2}
\notag\langle \bm S_y^2\rangle_\xi &=
\langle\alpha_H,0_V|
\bigl(\bm a_H^\dagger\bm a_V
+\bm a_H\bm a_V^\dagger\bigr)^2
|\alpha_H,0_V\rangle\\
\notag&=
\langle\alpha_H,0_V|
\bigl(
	\bm a_H^\dagger\bm a_V
	\bm a_H^\dagger\bm a_V
	+ \bm a_H^\dagger\bm a_V
	\bm a_H\bm a_V^\dagger\\
\notag &\hspace{1cm}	
	+\bm a_H\bm a_V^\dagger
	\bm a_H^\dagger\bm a_V
	+\bm a_H\bm a_V^\dagger
	\bm a_H\bm a_V^\dagger
\bigr)
|\alpha_H,0_V\rangle\\
&=|\alpha|^2, 
\end{align}
and 
\begin{align}\label{app:eq:Sx2}
\notag 
\langle \bm S_x^2\rangle_\xi &=
%\langle\alpha_x,0_y|
\langle\alpha_H,0_V|
\bigl(\bm a_x^\dagger\bm a_x
-\bm a_y^\dagger\bm a_y\bigr)^2
%|\alpha_x,0_y\rangle
|\alpha_H,0_V\rangle
\\
& = |\alpha|^2 + |\alpha|^4.
\end{align}

Then, we obtain the square error:
\begin{align}{\label{app:eq:err}}
\notag \epsilon_{\bm \sigma_z}^2 = 
\langle\bm N_{\bm \sigma_z}^2\rangle
&= \dfrac{1}{|\alpha|^2}
\big(\cot^2(2g) + 1\big)\\
&= \dfrac{1}{|\alpha|^2\sin^2(2g)}.
\end{align}

%Without loss of generality,
%we can consider the interaction is weak,
%such that $g \ll 1$, then 
%$\cot^2(2g)\approx 1/4g^2$.
%As a result, we obtain
%$\epsilon_{\bm \sigma_z}^2 
%\approx \frac{1}{4g^2|\alpha|^2} + 
%\frac{1}{|\alpha|^2}$.
%As  $g \ll 1$, we have 
%$\frac{1}{4g^2|\alpha|^2} \ll
%\frac{1}{|\alpha|^2}$
%and thus we can neglect 
%$\frac{1}{|\alpha|^2}$.
%Then, let $g|\alpha| = \chi$,
%the measurement strength,
%we obtain 
%$\epsilon_{\bm \sigma_z}^2 
%\approx \frac{1}{4\chi^2}$.
%as shown in Eq.~\eqref{eq:err}
%in the main text.

\subsection{The disturbance}

Next, we calculate the 
%{\it ms} 
{\it rms} 
disturbance,
Starting from $\bm B_T$ operator 
in Eq.~\eqref{app:eq:sigmaxf},
\begin{align}\label{app:eq:sigmaxfBT}
\notag \bm B_T &\equiv 
(\bm \sigma_x\otimes \bm I)_T\\
&= 
%\bm \sigma_x\otimes
%\cos(2g\bm S_z)
%- \bm \sigma_y\otimes
%\sin(2g\bm S_z). 
\bigl(
\bm \sigma_x\otimes
\cos(2g\bm S_z)
\bigr)_0
- 
\bigl(
\bm \sigma_y\otimes
\sin(2g\bm S_z)
\bigr)_0 .
\end{align}
The disturbance operator reads
\begin{align}
\bm D_{\bm \sigma_x} = 
\bigl(\bm \sigma_x\otimes\bm I\bigr)_T  
-\bigl(\bm \sigma_x\otimes\bm I\bigr)_0\;. \label{app:eq:dist}
\end{align}
Substituting Eq.~\eqref{app:eq:sigmaxfBT}
into Eq.~\eqref{app:eq:dist}
and taking the square of both sides, 
we have
\begin{align}
\bm D_{\bm \sigma_x}^2 = 
\Bigl[\underbrace{
\bigl(
	\bm \sigma_x\otimes
	\bigl[\cos(2g\bm S_z) - \bm I\bigr]
\bigr)_0
        }_{\bm X}-
        \underbrace{
\bigl(
	\bm \sigma_y\otimes\sin(2g\bm S_z)
\bigr)_0
        }_{\bm Y}
 \Bigr]^2.
\end{align}

Now, we calculate the average 
$\langle\bm D_{\bm \sigma_x}^2\rangle$ 
over the initial joint 
state $|\psi\rangle\otimes|\xi\rangle$.
We have
\begin{align}
\notag & \langle \bm X^2\rangle = 
\bigl\langle \bigl(\cos(2g\bm S_z) - \bm I\bigr)^2
 \bigr\rangle_\xi,\\
\notag & \langle \bm Y^2\rangle 
=
\bigl\langle \sin^2(2g\bm S_z)
\bigr\rangle_\xi,\\
\notag & \langle \bm X\bm Y\rangle 
=\langle \bm Y\bm X\rangle = 0.
\end{align}
Finally, we have 
\begin{align}\label{app:eq:D2}
\langle\bm D_{\bm \sigma_x}^2\rangle
= 2\bigl(1 -
\langle \cos(2g\bm S_z)\rangle_\xi\bigr).
\end{align}
For 
%$|\xi\rangle = |\alpha_x,0_y\rangle$
$|\xi\rangle = |\alpha_H,0_V\rangle$
and using the operator ordering relation
\cite{PhysRevD.35.1831,PhysRevA.38.2233,Fan_2003},
such as $e^{\kappa\bm a^\dagger\bm a}
= :e^{(e^\kappa-1)\bm a^\dagger\bm a}:$,
we have
\begin{align}\label{app:eq:j1}
\langle \cos(2g\bm S_z)\rangle_\xi
%= e^{-2|a|^2\sin^2(g)}.
= e^{-2|a|^2\sin^2g}.
\end{align}
Finally, we obtain the square disturbance:
\begin{align}{\label{app:eq:dis}}
\eta^2_{\bm \sigma_x} = 
\langle\bm D_{\bm \sigma_x}^2\rangle
%= 2(1-e^{-2|a|^2\sin^2(g)}),
= 2(1-e^{-2|a|^2\sin^2g}),
\end{align}
%where $\chi = g|\alpha|$.
%For small $g$, we obtain 
%$\eta^2_{\bm \sigma_x} 
%\approx 4\chi^2$,
as shown in Eq.~\eqref{eq:dis}
in the main text.

\setcounter{equation}{0}
\renewcommand{\theequation}{C.\arabic{equation}}
%\section{Class of polarization squeezed 
%coherent meter state}
\section{Phase-space 
approximation
and polarization-squeezed light meter}
\label{appC}

In this Appendix, we examine 
a class of the polarization squeezed 
coherent meter states
using phase-space approximation.
Let us consider a class of the 
squeezed coherent state of 
the two polarization modes as
\begin{align}\label{eq:appxisqueezed}
|\xi\rangle = |\alpha,z\rangle = 
\mathcal{D}(\alpha)
\mathcal{S}(z)
%|0_H0_V\rangle,
|0\rangle_H |0\rangle_V,
\end{align}
where $\mathcal{D}(\alpha) = 
{\rm exp}\bigl[\alpha \bm a_H^\dagger
-\alpha^*\bm a_H\bigr]$ 
is the displacement operator
with $\alpha = |\alpha| e^{i\phi}$ in the polar form,
and the two-mode squeezing operator is 
chosen to be
$\mathcal{S}(z) = {\rm exp}\bigl[z^*\bm a_L\bm a_R
-z\bm a_L^\dagger\bm a_R^\dagger\bigr]$
with $z = re^{i\vartheta}$, and $\bm a_{L(R)} =
\bigl(\bm a_H \pm i\bm a_V\bigr)/\sqrt{2}$.
This squeezing operator is 
%equivalence 
equivalent
to 
$ 
\mathcal{S}(z) 
=\mathcal{S}_H(z)\mathcal{S}_V(z) 
$
where
$
\mathcal{S}_H(z)=
{\rm exp}\bigl[\frac{1}{2}z^*\bm a_H^2
-\frac{1}{2}z(\bm a_H^\dagger)^2\bigr]
$
and
$
\mathcal{S}_V(z)=
{\rm exp}\bigl[\frac{1}{2}z^*\bm a_V^2
-\frac{1}{2}z(\bm a_V^\dagger)^2\bigr]
$.

%It is defined that when 
It is known that when 
$r>0$,
$\phi - \vartheta/2 = 0$
results in the amplitute-squeezed coherent state, while
the phase-squeezed coherent state
will happen when $\phi - \vartheta/2 = \pm \pi/2$
\cite{gerry_knight_2008}. 
%
%For simplicity, 
%demonstration, 
%we choose $\phi = 0$, i.e., 
%%$|\alpha|$ is a real number,
%\textcolor{red}{%
%$\alpha$ is a positive number,
%}%
%%then $\vartheta = 0\ (\pi)$ for  
%%amplitude (phase)-squeezed coherent state.
%and $\vartheta = 0$.  
In the following, we choose 
%$\vartheta = 2\phi$. 
$\phi - \vartheta/2 = 0$.
%so that
%the amplitude-squeezed coherent state 
%appears in the $H$-polarization mode for $r>0$.

The mean photon number in the two modes are 
\begin{align}\label{eq:appnxny}
\notag \langle n_H\rangle 
& 
= \langle \bm a_H^\dagger 
\bm a_H\rangle \\
\notag&
= \!_H\!\langle 0|\mathcal{S}_H^\dagger(z)
%\mathcal{D}_H^\dagger(\alpha)
\mathcal{D}^\dagger(\alpha)
\bm a_H^\dagger 
\bm a_H
%\mathcal{D}_H(\alpha)
\mathcal{D}(\alpha)
\mathcal{S}_H(z)|0\rangle_H
 \\
&= |\alpha|^2+{\rm sinh}^2r, \\
\langle n_V\rangle
&= 
\langle \bm a_V^\dagger 
\bm a_V\rangle
= {\rm sinh}^2r,
\end{align}
and the variances are
\begin{align}\label{eq:appvarinxny}
\sigma^2_{n_H} 
&=|\alpha|^2e^{-2r}+2\cosh^2r\sinh^2r \nonumber \\
&=|\alpha|^2e^{-2r}+\frac12\sinh^22r ,
\\
\sigma^2_{n_V} 
&=  \frac12\sinh^22r.
\end{align}
The
expectation values 
of the Stokes operators
give
\begin{align}\label{eq:appns}
\langle \bm S_0\rangle
&= |\alpha|^2+2{\rm sinh}^2r\;,\\
\langle \bm S_x\rangle &= |\alpha|^2,
\text{ and }
\langle \bm S_y\rangle = 
\langle \bm S_z\rangle = 0\;.
\end{align}
%
%\textcolor{blue}{
%Here, we have used
%useful transformations
%\begin{align}\label{eq:appuseful}
%\mathcal{D}^\dagger(\alpha)\bm a 
%\mathcal{D}(\alpha)
%&= \bm a + \alpha,\\
%\mathcal{D}^\dagger(\alpha)\bm a^\dagger 
%\mathcal{D}(\alpha)
%&= \bm a^\dagger + \alpha^*,\\
%
%\mathcal{S}^\dagger(z)\bm a 
%\mathcal{S}(z)
%&= \bm a\cosh r - \bm a^\dagger\sinh r,\\
%\mathcal{S}^\dagger(z)\bm a^\dagger
%\mathcal{S}(z)
%&= \bm a^\dagger\cosh r - \bm a\sinh r.
%\end{align}
%}
%
The variances give
\begin{align}\label{eq:appvariS0}
\sigma^2_{\bm S_0} = \sigma^2_{\bm S_x}=\sigma^2_{\bm S_y}
&=|\alpha|^2e^{-2r}+\sinh^22r 
\\
%\end{align}
%
%\begin{align}\label{eq:appSySz}
%\sigma^2_{\bm S_y} &= 
%|\alpha|^2e^{-2r}+{\rm sinh}^22r\quad \text{and }
\sigma^2_{\bm S_z} 
&=  |\alpha|^2e^{2r}.
\end{align}
For 
%$\alpha \gg {\rm sinh}\ r$, 
$|\alpha| \gg e^{3r}$, 
we can ignore the term 
${\rm sinh}^22r$ %in \eqref{eq:appvari}.
and read
\begin{align}\label{eq:appvariSall}
\sigma^2_{\bm S_0} = \sigma^2_{\bm S_x} =  
\sigma^2_{\bm S_y} &=|\alpha|^2e^{-2r},\\
%\sigma^2_{\bm S_y} &=  |\alpha|^2e^{-2r}, \quad \text{and }
\sigma^2_{\bm S_z} &= |\alpha|^2e^{2r}.
\end{align}
Thus, for $r>0$, we observe squeezing in 
$\sigma^2_{\bm S_0}$, $\sigma^2_{\bm S_x}$ and $\sigma^2_{\bm S_y}$,
while anti-squeezing appears in $\sigma^2_{\bm S_z}$.  

%For simplify, 
We 
%also 
introduce the
canonical operators
\begin{align}\label{eq:appxp}
\bm q = \dfrac{\bm S_y}
{\sqrt{|\langle\bm S_x\rangle|}}\; ,
\text{ and } 
%=\dfrac{1}{\sqrt{|\langle\bm S_x\rangle|}}
%\Bigl(\bm a_H^\dagger\bm a_V
%+\bm a_H\bm a_V^\dagger\Bigr),\\
\bm p = \dfrac{\bm S_z}
{\sqrt{|\langle\bm S_x\rangle|}}
%&=\dfrac{-i}{\sqrt{|\langle\bm S_x\rangle|}}
%\Bigl(\bm a_H^\dagger\bm a_V
%-\bm a_H\bm a_V^\dagger\Bigr),
\end{align}
which proportional to the polarized
Stokes operators $\bm S_y$
and $\bm S_z$, respectively.
The commutation relation
$[\bm q, \bm p] = 
2i{\bm S_x}/
{|\langle\bm S_x\rangle|}\simeq 2i$
indicates that
$\bm q$ and $\bm p$ can be regarded as a pair of canonical operators
when $\bm S_x$ can be approximated by a classical positive constant as $|\langle\bm S_x\rangle|$
so that 
$[\bm S_x, \bm S_y]\simeq 0$ and $[\bm S_x, \bm S_z]\simeq 0$.
It means that the third and higher terms in \eqref{app:eq:BCHE} can be ignored
as in the case 
%of the weak interaction approximation 
where $g\ll1$,
and that the Stokes operators, which essentially hold the discrete nature of the photon number, 
are replaced by the canonical continuous variable operators. 
Again, this approximation is valid only when 
$g\ll1$ so that the change in $\bm S_x/{|\langle\bm S_x\rangle|}$, 
which is in the order of $g^2$, is sufficiently small.

Then, 
the variances in $\bm q$ and $\bm p$ 
read
\begin{align} \label{eq:appvaris}
%\sigma_{\bm S_0} = \sigma_{\bm S_x} =
\sigma^2_{\bm q} =  e^{-2r}\ ;\
\sigma^2_{\bm p} = e^{2r}.
\end{align}
%
%Then, 
We can define
%a class of 
the wave function for
the polarization squeezed 
coherent state 
%wave function 
as
\begin{align}\label{eq:appsq}
\psi(q)= \Bigl(\dfrac{1}
{2\pi\sigma^2}\Bigr)^{1/4}
e^{-\frac{q^2}{4\sigma^2}},
\end{align}
where 
%$\sigma = e^{-r}$ 
$\sigma = \sqrt{\sigma_{\bm q}^2}= e^{-r}$ 
represents 
a squeezing parameter: 
\[ 
  \begin{cases}
    \sigma < 1 \to r > 0: & \quad \text{amplitude-squeezed}\\
     \sigma = 1 \to r = 0:
      & \quad \text{no squeezed}\\ 
    \sigma > 1 \to r < 0: & \quad \text{phase-squeezed}
  \end{cases}
\]
Then, the meter light state $|\xi\rangle$
can be defied as
\begin{align}\label{eq:appxid}
\notag|\xi\rangle &=\int \psi(q)
|q\rangle {\rm d}q\\
&=\Bigl(\dfrac{1}
{2\pi\sigma^2}\Bigr)^{1/4}
\int e^{-\frac{q^2}{4\sigma^2}}
|q\rangle{\rm d}q,
\end{align}
which we name as polarization squeezed
coherent state.

\subsection{The error}\label{app:C1}
The interaction evolution is defined by
\begin{align}\label{eq:uere}
\bm U_T = e^{-ig|\alpha|\bm \sigma_z\otimes \bm p},
\end{align}
Using the BCH formula, we have
\begin{align}\label{eq:app:sxtgauss}
\notag (\bm I\otimes \bm q)_T &= 
e^{ig|\alpha|\bm \sigma_z\otimes \bm p}\
(\bm I\otimes \bm q)_0\
e^{-ig|\alpha|\bm \sigma_z\otimes \bm p}\\
&=
(\bm I\otimes \bm q)_0 
+ 2g|\alpha|(\bm \sigma_z\otimes \bm I)_0.
\end{align}
Therefore, if we measure 
the calibrated meter operator 
$(\bm I\otimes\bm q)_T / 2g|\alpha|$,
we will obtain the corresponding 
information of $(\bm \sigma_z\otimes\bm I)_0$
in the system.

We first calculate the error
$\epsilon^2_{\bm \sigma_z} = 
\langle N^2_{\bm \sigma_z}\rangle_\xi = 
\langle\bm q^2\rangle_\xi/ 4g^2|\alpha|^2$.
Particularly, for 
$|\xi\rangle = 
\Bigl(\dfrac{1}{2\pi\sigma^2}\Bigr)^{1/4}\int 
e^{-\frac{q^2}{4\sigma^2}} |q\rangle \ {\rm d}q
$,
we have
\begin{align}\label{app:eq:x2}
\notag \langle\bm q^2\rangle_\xi
& =\Bigl(\dfrac{1}{2\pi\sigma^2}\Bigr)^{1/4}
\int e^{-\frac{q^2}{4\sigma^2}}\langle q| {\rm d}q
\cdot \int q_1^2 |q_1\rangle\langle q_1| {\rm d}q_1 \cdot  \\
\notag &\hspace{3.5cm}
\Bigl(\dfrac{1}{2\pi\sigma^2}\Bigr)^{1/4}
\int e^{-\frac{q_2^2}{4\sigma^2}} |q_2\rangle {\rm d}q_2\\
\notag &= 
\Bigl(\dfrac{1}{2\pi\sigma^2}\Bigr)^{1/2}
\int q^2 e^{-\frac{q^2}{2\sigma^2}}{\rm d}q\\
\notag & =
\Bigl(\dfrac{1}{2\pi\sigma^2}\Bigr)^{1/2}
\dfrac{1}{2}\sqrt{8\pi\sigma^6}\\
& = \sigma^2
\end{align}

Then, we obtain the square error
\begin{align}\label{eq:appse}
\epsilon_{\bm \sigma_z}^2
 =\dfrac{\langle\bm q^2\rangle_\xi}{4g^2|\alpha|^2}
 = \dfrac{1}{4\chi^2},
\end{align}
where in the last equally,
we have set $\chi = g|\alpha|/\sigma$.

\subsection{The disturbance}
%We next calculate the disturbance. 
Doing similar to App.~\ref{appB},
we obtain 
\begin{align}\label{app:eq:D2TT}
\langle\bm D_{\bm \sigma_x}^2\rangle
= 2\bigl(1 -
\langle \cos(2g|\alpha|\bm p)\rangle_\xi\bigr).
\end{align}
Using the Fourier transformation, 
we recast the meter state $|\xi\rangle$
in the momentum representation as 
$|\xi\rangle = 
\Bigl(\dfrac{2\sigma^2}{\pi}\Bigr)^{1/4}\int 
e^{\frac{-\sigma^2p^2}{4}} |p\rangle \ {\rm d}p$.
Then, we obtain
\begin{align}\label{app:eq:dis2}
\eta^2_{\bm \sigma_x} = 
\langle\bm D_{\bm \sigma_x}^2\rangle
= 2\bigl(1 -
e^{-2\chi^2}\bigr),
\end{align}
as shown in Eq.~\eqref{eq:ers}
in the main text.

\setcounter{equation}{0}
\renewcommand{\theequation}{D.\arabic{equation}}
\section{Weak Interaction Approximation (WIA)}
\label{appBS}
%We first provide that, 
Under the WIA, 
%ie.,
%i.e.,
we assume
$g \ll 1$
and
$g|\alpha| \ll 1$. 
%then 
From Eqs. (\ref{app:eq:err}, \ref{app:eq:dis}), 
we obtain
\begin{align}\label{appeq:errdis}
\epsilon_{\bm \sigma_z}^2 
\approx \frac{1}{4g^2|\alpha|^2}, \ \text{ and } \
\eta^2_{\bm \sigma_x} 
\approx 4g^2|\alpha|^2.
\end{align}
%
%The same relations are obtained
From Eqs. (\ref{eq:appse}, \ref{app:eq:dis2}),
we obtain for $\chi \ll 1$, 
\begin{align}\label{appeq:errdis2}
\epsilon_{\bm \sigma_z}^2 
\approx \frac{1}{4\chi^2}, \ \text{ and } \
\eta^2_{\bm \sigma_x} 
\approx 4\chi^2,
\end{align}
which are equivalent to Eqs.~\eqref{appeq:errdis} when $\sigma=1$.
Note that, in Eq.~\eqref{eq:appse},
we have already assumed that $g\ll1$ in the phase-space approximation.
%
%As a result, the 
In Eqs.~\eqref{appeq:errdis} and \eqref{appeq:errdis2},
we observe
$\epsilon_{\bm \sigma_z}^2
\eta_{\bm \sigma_x}^2=1$
and thus
Heisenberg-Arthurs-Kelly 
uncertainty is valid with minimal uncertainty.

%Next, 
We also show that,
under the WIA,
%we also provide 
%that in this case, 
both the noise and disturbance are unbiased. 
As described in the main text,
the noise operator 
\eqref{appeq:N} is 
already unbiased, i.e., 
$\langle\bm N_{\sigma_z}\rangle_\xi=0$ 
and thus $\langle\bm N_{\sigma_z}\rangle=0$ 
irrespective of $|\psi\rangle$,
provided that 
$\langle \bm S_y\rangle_\xi =0$.
Obviously, it is also true in the case of WIA.
%
%\eqref{appeq:N} is 
%recast as
%\begin{align}\label{appeq:Ns}
%\bm N_{\bm \sigma_z} 
% =  \dfrac{\bigl(\bm I\otimes \bm S_y\bigr)_0}
%{2g\langle\bm S_x\rangle_\xi}
%+\dfrac{(\bm \sigma_z\otimes \bm S_x)_0}
%{\langle\bm S_x\rangle_\xi}
% - \bigl(\bm \sigma_z\otimes\bm I\bigr)_0\;,
%\end{align}
%
%The average of noise operator is given by
%\begin{align}\label{appeq:AveN}
%\langle \bm N_{\bm \sigma_z} \rangle 
% =  \dfrac{\langle \bm S_y\rangle_\xi}
%{2g\langle\bm S_x\rangle_\xi}
%+\dfrac{\langle \bm \sigma_z\rangle_\psi \
%\langle \bm S_x\rangle_\xi}
%{\langle\bm S_x\rangle_\xi}
%-\langle \bm \sigma_z\rangle_\psi,
%\end{align}
%
%and we have 
%\begin{align}\label{app:eq:Syw}
%\langle \bm S_y\rangle_\xi =
%\langle\alpha_H,0_V|
%\bigl(\bm a_H^\dagger\bm a_V
%+\bm a_H\bm a_V^\dagger\bigr)
%|\alpha_H,0_V\rangle
%&=0,
%\end{align}
%then, 
%\textcolor{red}{%
%and thus
%}%
%we obtain
%\textcolor{red}{%
%$\langle \bm N_{\bm \sigma_z} \rangle_\xi =0$
%regardless of $|\psi\rangle$,
%given that 
%$\langle \bm S_y\rangle_\xi =0$.
%}%
%
%$\langle \bm N_{\bm \sigma_z} \rangle =0$
%, i.e., the noise operator is
%unbiased. 
%
%Similarly, 
For the 
%average of 
disturbance operator, 
from Eq.~\eqref{app:eq:sigmaxfBT}
and \eqref{app:eq:dist}, 
we get
%we have
%\begin{align}\label{appeq:BT}
%\notag 
%\bm B_T 
%%& = \bigl(\bm \sigma_x\otimes\bm I\bigr)_T \\
%& \approx \bigl(\bm \sigma_x\otimes\bm I\bigr)_0
%-2g \bigl(\bm\sigma_y\otimes \bm S_z\bigr)_0,
%\end{align}
%\textcolor{red}{%
%when $(g\bm S_z)^2$ and higher terms can be neglected.
%}%
%Then 
%disturbance operator is given by
\begin{align}\label{appeq:dist}
% \bm D_{\bm \sigma_x} 
%= 
%\bm B_T  
%-\bigl(\bm \sigma_x\otimes\bm I\bigr)_0
%= -2g \bigl(\bm\sigma_y\otimes \bm S_z\bigr)_0\;,
\bm D_{\bm \sigma_x} = 
\bigl(
	\bm \sigma_x\otimes
	\bigl[\cos(2g\bm S_z) - \bm I\bigr]
\bigr)_0
-\bigl(
	\bm \sigma_y\otimes\sin(2g\bm S_z)
\bigr)_0
\end{align}
and %the average gives
\begin{align}
\label{appeq:AvrDe}
%\langle\bm D_{\bm \sigma_x}\rangle = 
%-2g \langle\sigma_y\rangle_\psi
%\langle \bm S_z\rangle_\xi = 0,
\langle\bm D_{\bm \sigma_x}\rangle_\xi 
&= 
%-2g \bm\sigma_y
%\langle \bm S_z\rangle_\xi = 0,
	\bm \sigma_x %\otimes
\bigl[
\left\langle
	\cos(2g\bm S_z) \right\rangle_\xi - 1 \bigr]
 -\bm \sigma_y %\otimes
	\left\langle
\sin(2g\bm S_z)
\right\rangle_\xi
 \notag \\
 & = \bm \sigma_x \big( e^{-2|a|^2\sin^2g}-1 \big).
% \approx 0
\end{align}
%where
Here, we use Eq.~\eqref{app:eq:j1} and
$\langle \sin(2g\bm S_z)\rangle_\xi=0$.
Thus, $\langle\bm D_{\bm \sigma_x}\rangle_\xi\approx0$
%provided that
%$ \langle \bm S_z\rangle_\xi 
%= -i\langle\alpha_H,0_V|
%\bigl(\bm a_H^\dagger\bm a_V
%-\bm a_H\bm a_V^\dagger\bigr)
%|\alpha_H,0_V\rangle
%= 0$, 
%
%$2g^2\langle\bm S_z^2\rangle_\xi\ll1$ 
%and higher order terms of
%$\langle\bm S_z^k\rangle_\xi$ 
%are negligible.
%In our case, 
%where $ \langle \bm S_z\rangle_\xi=0$ and 
%$\langle\bm S_z^2\rangle_\xi=|\alpha|^2$,
%the conditions are satisfied 
when 
$g\ll1$ and $g|\alpha|\ll1$.
Consequently,
for our initial meter state 
under the WIA,
%$|\xi\rangle$ 
%where 
%$\langle \bm S_y\rangle_\xi =
%\langle \bm S_z\rangle_\xi = 0$,
%it is obvious that
%Obviously, 
both the noise and disturbance operators
are unbiased,
i.e., 
 $\langle\bm N_{\sigma_z}\rangle =
 \langle\bm D_{\sigma_x}\rangle =0$ 
irrespective of the initial system state $|\psi\rangle$.
This {\it joint-unbiassedness} condition is
%which is 
sufficient 
for holding the 
Heisenberg-Arthurs-Kelly uncertainty
\cite{PhysRevA.67.042105}.

\bibliography{refs}
\end{document}